\documentstyle[aps,preprint]{revtex}
\begin{document}
\draft
\tighten
\preprint{\vbox{
                \hfill TIT/HEP-428/NP 
 }}
\title{Meson-baryon couplings and the $F/D$ ratio from QCD sum rules}
\author{Hungchong Kim$^1$ \footnote{E-mail : 
hckim@th.phys.titech.ac.jp, JSPS fellow},
Takumi Doi and
Makoto Oka }
\address{Department of Physics, Tokyo Institute of Technology, Tokyo 
152-8551, 
Japan }
\author{Su Houng Lee}
\address{Department of Physics, Yonsei University, Seoul 120-749, Korea}

\maketitle
\begin{abstract}
Motivated by the recent work on the calculation of the $\pi NN$ coupling 
constant using QCD sum rule beyond  the chiral limit,
we construct the corresponding sum rules for the 
couplings, $\eta NN$, $\pi \Xi\Xi$, $\eta \Xi\Xi$,
$\pi \Sigma \Sigma$ and $\eta \Sigma \Sigma$.
In constructing the $\eta$-baryon sum rules, 
we use the second moment of the $\eta$ wave function,  which we obtain  
from the pion wave function after SU(3) rotation.   
In the  SU(3) symmetric limit, we can identify the term responsible for the 
$F/D$ ratio in the OPE, which after the sum rule analysis
 gives  $F/D\sim 0.2$, a factor of 3 smaller
than  from other studies.
We also present a qualitative analysis including the  
 SU(3) breaking terms.

\end{abstract}

\section{INTRODUCTION}
\label{sec:intro}

Meson-baryon couplings in the SU(3) sector 
are important quantities to be determined in
modern nuclear physics.  Each coupling, representing the strength of
baryon-baryon interaction to each mesonic channel, is an important
ingredient for a systematic analysis of baryon-baryon scatterings and 
meson productions off a baryon.  
Experimentally, the couplings are determined by fitting the
experimental data of meson-baryon scatterings, baryon-baryon phase
shift analysis~\cite{bonn,rijken}.   However, in general, it is difficult 
to pin down a specific channel from a certain reaction and
determine unambiguously the coupling of concern.  In addition, hadronic
models used in fitting processes are not often unique, which provides 
an additional limitation of the conventional methods.  

SU(3) symmetry, as it provides a systematic classification of mesons
and baryons, is expected to govern the meson-baryon couplings. Indeed,
according to Ref.~\cite{swart}, all the meson-baryon couplings in the
SU(3) limit satisfy simple relations represented by the $\pi NN$ coupling, 
which is well-known experimentally,  
and the $F/D$ ratio.  This systematic classification of the
couplings is a basis for making realistic potential models for 
hyperon-baryon interactions~\cite{rijken}.  In this approach, the
$F/D$ ratio is an input for the analysis, usually
determined from other sources for example the SU(6) consideration.
However, a more realistic and self-contained method is to determine the 
ratio within the SU(3) classification.  One such method is to invoke 
QCD sum rules~\cite{SVZ} and calculate the meson-baryon couplings in 
the SU(3) limit.  This not only provides a QCD prediction for the $F/D$ ratio 
but also gives insights as to  how the couplings should be constrained 
even in  the SU(3) breaking case.

QCD sum rules~\cite{SVZ} utilize the two aspects of QCD, 
perturbative and nonperturbative effects, in 
representing hadronic spectral properties in terms of QCD parameters.
Within this framework, a correlation function of QCD operators is evaluated
via the operator product expansion (OPE).  This expansion includes the
perturbative part as well as the nonperturbative effects, which are 
systematically included by 
the power corrections.  In the hadronic side, an ansatz for the correlator 
is introduced in terms of hadronic degrees of freedom and
it is matched with the OPE.  The matching provides the hadronic 
parameter of concern in terms of QCD parameters.
QCD sum rules have been widely 
used~\cite{qsr} to predict baryonic masses, hadronic coupling 
constants and so on.  In some cases, they are known to be successful 
even though some cares must be taken in constructing a sum 
rule~\cite{hung2,hung4}.

When QCD sum rules are applied to meson-baryon couplings, a good 
place for testing QCD sum rules is to calculate the $\pi NN$ 
coupling $g_{\pi N}$~\cite{qsr,hung2,hat,hung1,hung3}
as its value is relatively well-known experimentally.  
A successful reproduction of this coupling within QCD sum rules 
may provide a solid framework to extend its method to other (not well-known)
meson-baryon couplings.
Indeed, in a series of papers~\cite{hung1,hung2,hung3},
the $\pi NN$ coupling has been calculated using the correlation 
function of the nucleon interpolation field $J_N$,
\begin{eqnarray}
\Pi (q, p) = i \int d^4 x e^{i q \cdot x} \langle 0 | T[J_N (x) 
{\bar J}_N (0)]| \pi (p) \rangle \ .
\label{two}
\end{eqnarray}
This two-point correlation function
with a pion seems to be more suitable than the three-point 
function approaches~\cite{yazaki} because we do not
need to worry about the contribution from the higher resonances 
$\pi(1300)$ and $\pi(1800)$
which could introduce substantial errors in determining the pion-nucleon 
coupling~\cite{maltman}. Moreover, 
using the two-point correlation function, the sum rule can be 
straightforwardly
extended to other meson-baryon
couplings and the SU(3) symmetric limit can be easily recovered.

One of us\cite{hung3} has recently proposed to construct a QCD sum rule 
starting from  Eq.~(\ref{two}) and going 
 beyond the chiral limit.  There~\cite{hung3}, a QCD sum rule
for the $i\gamma_5$ structure at the $p^2=m_\pi^2$ order is constructed.  
One interesting 
observation made in Ref.~\cite{hung3} is that 
the quark mass terms included as a consistent chiral counting
are found to be important in stabilizing
the sum rule and crucial in producing $\pi NN$ coupling close to
its empirical one.  
The uncertainties from QCD parameters in the extracted $g_{\pi N}$ are
estimated to be around $\pm 1$, sensitivity to the continuum 
threshold is found to be
small, and the unknown single-pole term which appears due to
the transition of $N\rightarrow N^*$ is also estimated to be
small.  Hence, this sum rule beyond the chiral limit
seems to have nice features and
may provide a reliable framework for extending to other meson-baryon
couplings.

In this work, we apply the QCD sum rules beyond the
chiral limit to other meson-baryon couplings in the
SU(3) sector.  By keeping the SU(3) symmetry in constructing the sum rules,
we can compare the OPE of each sum rule to the SU(3) relation of the 
coupling, which
allows us to identify the OPE terms generating the $F/D$ ratio.
Then a subsequent Borel analysis of the sum rule determines the
$F/D$ ratio.  Reliability of this value can be checked by 
further analyzing
other meson-baryon sum rules. 
Furthermore, 
the sum rules at the SU(3) limit will give us a 
hint how the SU(3) symmetry breaking is reflected in the couplings.
By identifying differences in the OPE from each sum rule, we might
be able to predict how each OPE term should be modified as 
we switch on the SU(3) breaking effect. Even though we do not precisely
know the OPE in the SU(3) breaking limit, we may
still see a general trend on how the physical couplings should be,
which can act as constraints to be satisfied in experimental analysis.

For this purpose, we reconstruct QCD sum rules for the $\pi NN$
coupling and extend the framework to $\eta NN$,
$\pi \Xi \Xi$, $\eta \Xi \Xi$, $\pi \Sigma \Sigma$,
$\eta \Sigma \Sigma$.
One of these extended sum rules will be compared to the 
$\pi NN$ sum rule
and used to estimate the $F/D$ ratio.  The other sum rules provide 
consistency checks for this value within the same framework. 
In the case of the $\eta$-baryon sum rules, we assume the
second moment of the twist-3 $\eta$ wave function to be the
same as the pion case. This assumption is supported by the
OPE satisfying the classification suggested from the SU(3)
symmetry.  We then speculate qualitatively how the SU(3) breaking
is reflected in the couplings.

The paper is organized as follows. In Section~\ref{sec:pnn},
we reconstruct the QCD sum rule for the $\pi NN$ coupling 
beyond the chiral limit.  
We apply the similar sum rule to the $\eta NN$ coupling in 
Section~\ref{sec:enn}.
We construct the sum rules for  
$\pi \Xi \Xi$ and $\eta \Xi \Xi$ in Section~\ref{sec:xisum}, 
$\pi \Sigma \Sigma$ and $\eta \Sigma \Sigma$ in  
Section~\ref{sec:sigmasum}.
In Section~\ref{sec:anal}, we present
our numerical analysis for the couplings in the SU(3) limit and
provide constraints for the $F/D$ ratio.
In Section~\ref{sec:anal2}, we qualitatively study how
the SU(3) breaking appears in the couplings.
We summarize in 
Section~\ref{sec:sum}.

\section{QCD sum rules for Pion-nucleon coupling}
\label{sec:pnn}

In this section, we construct a QCD sum rule for $\pi NN$ coupling beyond the
chiral limit.  The content of this part can be found in Ref.~\cite{hung3} but
we present this sum rule again with
some technical details that can be straightforwardly extended
for other meson-baryon couplings in later  sections.

We consider the two-point correlation function with
a pion,
\begin{eqnarray}
\Pi (q,p) = i \int d^4 x e^{i q \cdot x} \langle 0 | T[J_p (x) 
{\bar J}_p (0)]| \pi^0 (p) \rangle 
\equiv \int d^4 x e^{i q \cdot x} \Pi(x,p)\ .
\label{two2}
\end{eqnarray}
$J_p$ is the proton interpolating field suggested by Ioffe~\cite{ioffe1},
\begin{eqnarray}
J_p = \epsilon_{abc} [ u_a^T C \gamma_\mu u_b ] \gamma_5 \gamma^\mu d_c\ ,
\end{eqnarray}
where $a,b,c$ are color indices, $T$ denotes the transpose with
respect to the Dirac indices, $C$ the charge 
conjugation.  From this correlator, we collect the
terms contributing to the $i\gamma_5$ Dirac structure and expand
them in terms of the pion momentum $p^\mu$.  The correlator
in this expansion takes the form,
\begin{eqnarray}
\Pi_0 (q^2) + p\cdot q \Pi_1 (q^2) + p^2 \Pi_2 (q^2) + \cdot \cdot \cdot \ .
\end{eqnarray}
As the nucleon momentum $q_\mu$ is independent from the pion momentum
$p_\mu$, each scalar function can be used to construct a QCD sum rule.
The correlator at the soft-pion limit
$\Pi_0$ is equivalent to the nucleon chiral-odd sum 
rule~\cite{hat}: it does not provide an independent
determination of the coupling.  The OPE contributing to $\Pi_1$ is basically
the same as the $\Pi_0$ sum rule, again not useful
for calculating the coupling. 
We therefore consider $\Pi_2$ in constructing a sum rule.
By putting
the pion on its mass-shell $p^2=m_\pi^2$, we construct a sum rule
beyond the chiral limit,
which will provide a prediction for the pion-nucleon coupling
$g_{\pi N}$ beyond the chiral limit.  As we will see,
it is at this order that each meson-baryon sum rule is distinct
from the other sum rules.

In constructing the phenomenological side, we replace the nucleon interpolating
field with the physical nucleon field, $J_p \rightarrow \lambda_N \psi_p$,
and, using the pseudoscalar Lagrangian,
$g_{\pi N} {\bar \psi}_p i\gamma_5 \psi_p \pi^0$, we evaluate the
correlator in terms of hadronic degrees of freedom. 
At the chiral order $p^2=m_\pi^2$ from the correlator containing
the $i\gamma_5$ Dirac structure, 
the phenomenological correlator takes the form 
\begin{eqnarray}
m_\pi^2 \Pi^{phen}_2 (q^2) = 
m_\pi^2 {g_{\pi N} \lambda_N^2 \over (q^2-m^2_N)^2} + \cdot \cdot \cdot\ .
\label{pphen}
\end{eqnarray} 
The ellipses denote the contributions when $J_p$ couples to higher resonances.
This includes the continuum contribution whose spectral density
is parametrized by a step function with a certain threshold $S_0$ and
the single-pole terms associated with the transitions 
$N\rightarrow N^*$~\cite{ioffe2}.
Here $m_N$ denotes the nucleon mass.
 
In the OPE,  we only keep the quark-antiquark component of the 
pion wave function and use the vacuum saturation hypothesis
to factor out 
higher dimensional operators in terms of the pion wave function and the 
vacuum expectation values.  Accordingly, it is straightforward to
write the correlator in the coordinate space,
\begin{eqnarray}
\Pi(x,p) = -i \epsilon_{abc} \epsilon_{a'b'c'}
\Bigg \{&& \gamma_5 \gamma^\mu D^d_{cc'} \gamma^\nu \gamma_5
{\rm Tr} \left[i S_{aa'}(x)(\gamma_\nu C)^T iS_{bb'}^T(x) 
(C\gamma_\mu)^T \right ] \nonumber \\
&-&
\gamma_5 \gamma^\mu D^d_{cc'} \gamma^\nu \gamma_5
{\rm Tr} \left[i S_{ab'}(x)\gamma_\nu C iS_{ba'}^T(x) 
(C\gamma_\mu)^T \right ]\nonumber \\
&-&
\gamma_5 \gamma^\mu iS_{cc'}(x) \gamma^\nu \gamma_5
{\rm Tr} \left[i S_{ab'}(x)\gamma_\nu C (D^u_{ba'})^T    
(C\gamma_\mu)^T \right ]\nonumber \\
&+&
\gamma_5 \gamma^\mu iS_{cc'}(x) \gamma^\nu \gamma_5
{\rm Tr} \left[i S_{aa'}(x) (\gamma_\nu C)^T (D^u_{bb'})^T    
(C\gamma_\mu)^T \right ]\nonumber \\
&+&
\gamma_5 \gamma^\mu iS_{cc'}(x) \gamma^\nu \gamma_5
{\rm Tr} \left[ D^u_{aa'}(\gamma_\nu C)^T iS_{bb'}^T(x)    
(C\gamma_\mu)^T \right ]\nonumber \\
&-&
\gamma_5 \gamma^\mu iS_{cc'}(x) \gamma^\nu \gamma_5
{\rm Tr} \left[ D^u_{ab'}\gamma_\nu C iS_{ba'}^T(x)    
(C\gamma_\mu)^T \right ]
\Bigg \}\ .
\label{xp}
\end{eqnarray}
The quark propagators $iS(x)$ inside the
traces are the u-quark propagators and the ones outside of
the traces are the d-quark propagators.
Since we are interested in the $i\gamma_5$ Dirac structure, 
in most OPE we can replace the quark-antiquark component with a pion
as follows,
\begin{eqnarray}
(D_{aa'}^u)^{\alpha \beta} \equiv
\langle 0 | u^\alpha_a (x) {\bar u}^\beta_{a'} (0) | \pi^0 (p) \rangle\ 
\rightarrow
{\delta_{a a'} \over 12 } (i \gamma_5)^{\alpha \beta}
\langle 0 |
{\bar u}(0) i \gamma_5  u (x) | \pi^0 (p) \rangle\ .
\label{diquark}
\end{eqnarray}
A similar relation holds for the d-quark component.
Contributions from the other Dirac 
structures, $\gamma_5 \gamma_\mu$ and $\gamma_5 \sigma_{\mu\nu}$,
to this quark-antiquark component do not participate to the $i\gamma_5$ 
sum rule at the chiral order that we are considering.

We are concerned with the sum rule for the $i\gamma_5$ structure
at the order $p^2=m^2_\pi$. 
The $p_\mu$ dependence appears
only in the quark-antiquark component.
Obviously,  
the second order terms in the expansion of the pion matrix element
in $p_\mu$ should contribute to the sum rule.
In a consistent approach at this order, 
terms linear in quark mass ($m_q$) should also be included 
as $m_q$ is the same chiral
order with $m_\pi^2$
via the Gell-Mann$-$Oakes$-$Renner relation,  
\begin{eqnarray}
-2 m_q \langle {\bar q} q\rangle = m_\pi^2 f_\pi^2\ .
\end{eqnarray}
These terms  are
obtained by taking $m_q$ terms from a quark propagator and at the
same time taking the zeroth moment of the pion matrix element.
Additional contributions at this order are from the three-particle
wave function, obtained by taking a gluon tensor from a quark propagator
and moving into the quark-antiquark component,
\begin{eqnarray}
\langle 0 |g_s G_{\mu\nu}^A (0) d^\alpha_a (x)
{\bar d}^\beta_b (0) |\pi^0 (p)\rangle
\rightarrow  {i f_{3\pi} \over 32 \sqrt{2} } m_\pi^2 t^A_{ab}
(\gamma_5 \sigma_{\mu\nu})^{\alpha \beta}\ ,
\label{gluon}
\end{eqnarray}
where
$f_{3\pi}=0.003$ GeV$^2$~\cite{bely}
and the color matrices $t^A$ are normalized $tr(t^A t^B)=\delta^{AB}/2$.

With these in mind, we calculate the correlator 
Eq.~(\ref{xp}) using the
quark propagator given in Ref.~\cite{wilson}\footnote{Note, 
the gluon strength tensor used in Ref.~\cite{wilson} has opposite sign from
the one in Ref.~\cite{qsr}. This difference is simply due to how
the covariant derivative is defined.}. 
The OPE up to dimension 8 in the coordinate space is 
\begin{eqnarray}
&+&{2 i \over \pi^4} 
\langle 0 | {\bar d} (0) i \gamma_5 d(x) |\pi^0 \rangle 
{1\over x^6}
-
{3if_{3\pi}m^2_\pi \over 4 \sqrt{2} \pi^4} {1\over x^4} \nonumber \\
&-&
{i \over 96 \pi^2} \left \langle {\alpha_s \over \pi} {\cal G}^2 
\right \rangle \langle 0 | {\bar d} (0) i \gamma_5 d(x) |\pi^0 \rangle
{1\over x^2} 
-
{i \over 2 \pi^2} m_u \langle {\bar u} u \rangle 
\langle 0 | {\bar d} (0) i \gamma_5 d(x) |\pi^0 \rangle {1\over x^2}
\nonumber \\
&+&
{i m_u \langle {\bar d} g_s \sigma \cdot {\cal G} d \rangle
+i m_d \langle {\bar u} g_s \sigma \cdot {\cal G} u 
\rangle \over 48 \pi^2}
\langle 0 | {\bar u} (0) i \gamma_5 u(x) |\pi^0  \rangle 
\left [ ln({-x^2 \Lambda^2_c \over 4}) + 2 \gamma_{EM} \right ]\ .
\label{pope}
\end{eqnarray}
Except for the last term containing the quark-gluon mixed condensate,
all others come from the first two terms in Eq.~(\ref{xp}).
The other four terms in Eq.~(\ref{xp}), as they have the 
quark-antiquark component inside of 
the traces, contribute mostly zero to the OPE except producing the
last term in Eq.~(\ref{pope}).  Note, only the last term contains 
the u-quark component 
with a pion while the others contain the d-quark component.
This identification will be useful for later developments.
Furthermore, as the last term
contains the quark mass $m_q$ which is the chiral
order that we are considering,  
we need to
take only the local contribution to the u-quark component with a pion 
(the part that survives in the soft-pion limit.).
Under the Fourier transformation to the
momentum space,
the cutoff mass $\Lambda_c$ and Euler-Mascheroni constant $\gamma_{EM}$
will disappear.

The pion matrix element appearing in the OPE
can be written in terms of the twist-3 pion wave function,
\begin{eqnarray}
\langle 0 |
{\bar u}(0) i \gamma_5  u (x) | \pi^0 (p) \rangle 
&=&-{\langle {\bar u} u\rangle \over f_\pi}\int^1_0 dt e^{-itp\cdot x} 
\varphi_p (t)\ ,\nonumber \\
\langle 0 |
{\bar d}(0) i \gamma_5  d (x) | \pi^0 (p) \rangle 
&=& +{\langle {\bar d} d\rangle \over f_\pi}\int^1_0 dt e^{-itp\cdot x} 
\varphi_p (t)\ .
\label{twist3}
\end{eqnarray}
The zeroth and the second moment of this twist-3 pion wave function
are needed in constructing our sum rule.
Note that the overall normalization of the matrix element is fixed by
the soft-pion theorem, which gives opposite signs between the d-quark and
the u-quark component. 
Also, the soft-pion theorem fixes the zeroth moment of the pion wave 
function to $\int^1_0 dt \varphi_p (t)=1$.  
The twist-3 wave function is determined uniquely if the 
three-particle wave function is known~\cite{braun}. However,
the three-particle wave function gives only small corrections to the
asymptotic form of the twist-3 wave function [$\varphi_p (t) =1$].
Therefore, the second moment obtained from the asymptotic wave
function, which is fixed to $\int^1_0 dt~ t^2 
\varphi_p (t)=1/3$, is not so different from the moment using more realistic
wave function~\cite{bely,zhit}.

Using the zeroth moment of the wave function for the OPE containing $m_q$
and the second
moment for the rest of the OPE (except for the
term coming from the three-particle wave function), we obtain after
Fourier transformations,
\begin{eqnarray}
m_\pi^2 \Pi^{ope}_2 (q^2) &=&
m_\pi^2 ln(-q^2) \left [ {\langle {\bar q}q \rangle \over 12 \pi^2 f_\pi}
              + {3 f_{3\pi} \over 4\sqrt{2}\pi^2} \right ]
- 2 m_q \langle {\bar q}q \rangle^2  {1\over f_\pi q^2}
\nonumber \\
&+& m_\pi^2 {1\over 72 f_\pi} \langle {\bar q}q \rangle
\left \langle {\alpha_s \over \pi} {\cal G}^2
\right \rangle {1\over q^4}
+{2\over 3} m_q m_0^2 \langle {\bar q}q \rangle^2 
{1\over f_\pi q^4}\ .
\label{popep}
\end{eqnarray} 
Here the quark-gluon mixed condensate is parametrized as
$\langle {\bar d} g_s \sigma \cdot {\cal G} d \rangle \equiv
m_0^2 \langle {\bar d} d\rangle$  and similarly for the u-quark
with $m_0^2 =0.8$ GeV$^2$. 
In obtaining Eq.~(\ref{pope}), we have taken 
the isospin symmetric limit,
$\langle {\bar u}u \rangle=\langle {\bar d}d \rangle\equiv 
\langle {\bar q}q \rangle$ and $m_u =m_d \equiv m_q$.

We now match the OPE with its phenomenological counterpart [Eq.~(\ref{pphen})]
using  the single-variable dispersion 
relation~\cite{hung4}~\footnote{
In QCD sum rules with external fields, the double dispersion relation
has been proposed as a proper representation of the 
correlator~\cite{ioffe3}.
As the correlator in the tree level contains 
two baryonic propagators, it seems reasonable to use the
double dispersion relation.  This is however misleading because
the double dispersion relation
produces spurious terms coming from subtraction terms which
should not contribute to QCD sum rules.  This is because
the spectral density obtained from the double dispersion relation
is not compatible with the duality assumption used for modeling
the continuum contribution~\cite{hung4}.  When the spurious terms
are subtracted out from the sum rule using the double dispersion 
relation,  the resulting sum rule is equivalent to the one using
the single dispersion relation.}.
Under the Borel transformation with respect to $-q^2$, we obtain the
sum rule,
\begin{eqnarray}
&&g_{\pi N} m_\pi^2 \lambda^2_N e^{-m^2_N/M^2} [ 1+ AM^2]=
\nonumber \\
&&
-m_\pi^2 M^4 E_0 (x) \left [
{\langle {\bar q}q \rangle \over 12 \pi^2 f_\pi}
              + {3 f_{3\pi} \over 4\sqrt{2}\pi^2} \right ]
+ {2 m_q\over f_\pi} \langle {\bar q}q \rangle^2 M^2
+ {m_\pi^2 \over 72 f_\pi} \langle {\bar q}q \rangle
\left \langle {\alpha_s \over \pi} {\cal G}^2\right \rangle
+{2m_q\over 3 f_\pi } m_0^2 \langle {\bar q}q \rangle^2\nonumber \\
&&\equiv {\cal O}_1 + {\cal O}_2 +{\cal O}_3+{\cal O}_4\ .
\label{sum1}
\end{eqnarray}
In the RHS, only  the last term comes from
the u-quark component with a pion, all others come  from the
d-quark component.   Note that ${\cal O}_2$ and ${\cal O}_4$ contain
the quark-mass.   They are included as a
consistent chiral counting.
The contribution from $N \rightarrow N^*$~\cite{ioffe2} is denoted by
the unknown constant $A$. 
The continuum contribution is included by 
the factor, $E_n(x \equiv S_0 /M^2)= 1- (1 + x +
\cdot\cdot \cdot + x^n/n!) e^{-x}$
where $S_0$ is the continuum threshold.  
The quark-mass dependence can be converted to  
the $m^2_\pi$ dependence via
the Gell-Mann$-$Oakes$-$Renner relation, which then
can be taken out from both sides as an overall factor.
This sum rule, when combined with the nucleon chiral odd sum rule,
gives a reliable value for the  the pion-nucleon coupling $g_{\pi N}$ 
with small uncertainty and  very closed to its
empirical value~\cite{hung3}.  

Before closing this section, we comment about the
sign of each contribution in the  OPE Eq.~(\ref{sum1}).
That is, ${\cal O}_1$, ${\cal O}_2$, ${\cal O}_4$
contribute with the same sign, while ${\cal O}_3$ contributes 
with the opposite sign.  However, the magnitude of 
${\cal O}_3$ is only  7\% of ${\cal O}_4$.  
Therefore, most OPE terms add up in producing the $\pi NN$ coupling.

\section{QCD sum rules for $\eta NN$ coupling}
\label{sec:enn}
 
Motivated by the $\pi NN$ sum rule beyond the chiral limit, 
we construct in this section a sum rule for the $\eta NN$ coupling.
As in  the $\pi NN$ sum rule calculation, we consider 
the $i\gamma_5$ structure from the correlator,
\begin{eqnarray}
 i \int d^4 x e^{i q \cdot x} \langle 0 | T[J_p (x) 
{\bar J}_p (0)]| \eta (p) \rangle \ ,
\label{etwo}
\end{eqnarray}
at the order $p^2=m^2_\eta$.
In the followings, we neglect the mixing 
between $\eta$-$\eta^\prime$ and therefore $\eta$ is assumed to be
$\eta_8$. This is exact in the SU(3) limit and it does not affect our
determination of the $F/D$ ratio below.

The phenomenological side can be constructed as before by
expanding the correlator  in $p_\mu$ and collecting terms proportional
to $p^2=m_\eta^2$.  The phenomenological side takes the
form
\begin{eqnarray}
m_\eta^2 {g_{\eta N} \lambda_N^2 \over (q^2-m_N^2)^2} + \cdot \cdot \cdot\ .
\label{ephen}
\end{eqnarray}
Note that the coupling $g_{\eta N}$ in this expansion is defined at 
the kinematical 
point $p^2=0$,
that is, $g_{\eta N} (p^2=0)$.
But the physical coupling is defined at $p^2=m_\eta^2$.
The correction of this kind is negligible in the $\pi NN$ case.
However, in the present case where
$m_\eta^2$ is much heavier than $m_\pi^2$, one might expect
some corrections from 
the form factor.  The monopole type form factor is often used for
meson-baryon couplings~\cite{bonn}.  In the present case, 
the coupling should be written
\begin{eqnarray}
g_{\eta N}(p^2) = g_{\eta N}
{\Lambda^2 -m_\eta^2 \over \Lambda^2 -p^2}\ ,
\end{eqnarray}
with $\Lambda \sim 1.5$ GeV according to Bonn potential~\cite{bonn}.
By expanding the form factor in $p^2$, 
we have
\begin{eqnarray}
{\Lambda^2 -m_\eta^2 \over \Lambda^2 -p^2}
= \left (1-{m_\eta^2 \over \Lambda^2} \right ) +
p^2 {\Lambda^2-m_\eta^2 \over \Lambda^4}+ \cdot \cdot \cdot\ .
\label{fexp}
\end{eqnarray}
The coupling $g_{\eta N}$ appeared in Eq.~(\ref{ephen})
is the physical coupling multiplied by the first term in this expansion,
which will be determined in this method.  The correction to $g_{\eta N}$
as we move from $p^2=0$ to $p^2=m_\eta^2$ is $m_\eta^2 /\Lambda^2\sim 0.13$.
Therefore, from this form factor effect,
the physical coupling should be larger by 13 \% than what we will 
determine in this work. 
Of course, more softer cut-off like $\Lambda 
\sim 1$ GeV will increase
the physical coupling slightly more and we will 
discuss the uncertainty coming from $\Lambda$  further in 
Section~\ref{sec:anal2}.
The second term in Eq.~(\ref{fexp}) is combined with
the correlator at the soft-meson limit to produce
a monopole structure [$\sim 1/(q^2-m_N^2)$], which together with
the unknown pole of $N\rightarrow N^*$ will be determined
via the best-fitting method.

The OPE is calculated similarly as before by factorizing the $\eta$ matrix 
element and the vacuum expectation values.
The quark-antiquark component with $\eta$ appearing in the OPE is
written similarly as the pion wave function.  Namely, the zeroth
moment of the wave function is determined by the soft-meson theorem but
the rest of the  $p_\mu$ dependence is absorbed into the $\eta$ wave function,
\begin{eqnarray}
\langle 0 |
{\bar u}(0) i \gamma_5  u (x) | \eta (p) \rangle 
&=&-{\langle {\bar u} u\rangle \over f_\eta \sqrt{3}}
\int^1_0 dt e^{-itp\cdot x} 
\varphi_{\eta} (t)\nonumber\ , \\
\langle 0 |
{\bar d}(0) i \gamma_5  d (x) | \eta (p) \rangle 
&=&-{\langle {\bar d} d\rangle \over f_\eta \sqrt{3}}
\int^1_0 dt e^{-itp\cdot x} 
\varphi_{\eta} (t)\ .
\label{etwist3}
\end{eqnarray}
We see a clear distinction from the pion case:
the u-quark and d-quark components with $\eta$ have the {\it same} overall
sign in contrast to Eq.~(\ref{twist3}). This is because $\eta$ is
an isoscalar particle. The phase convention in fixing the overall sign 
is consistent with the model by de Swart~\cite{swart}, which has been used in
constructing the phenomenological part.

In this definition, the zeroth moment is given by $\int^1_0 dt 
\varphi_{\eta} (t) =1$ as the coefficient is governed by the
soft-meson theorem.  The use of the soft-meson theorem should 
be fine in the SU(3) limit.  For the second moment, we take 
$\int^1_0 dt~t^2 \varphi_{\eta} (t) =1/3$ just like the pion
case.  This is certainly reasonable in the SU(3) limit
because in this limit the way of determining the second moment of 
the pion wave function~\cite{zhit} can be applied equally to the $\eta$
case.  This also makes sense as the OPE satisfies 
the SU(3) relation as we will see. A question remains as to how
one can model this second moment when the SU(3) symmetry is broken.
The second moment multiplied by $p^2=m_\eta^2$ will
contribute to our sum rule.
The physical mass of $\eta$ is much heavier than the pion. 
Under the same assumption for the second moment, the $\eta$ matrix
element will have an enhancing factor of $m_\eta^2/m_\pi^2\sim 15.6$
compared to the pion matrix element at the same chiral order.
This casts some  doubts on the parameterization in Eq.~(\ref{etwist3}).
This is a limitation of our current approach and needs to be 
improved in future. 
Nevertheless, in this work we will fix the second moment of 
the $\eta$ wave function as in the pion case, because the extra 
modification of the second moment does not affect strongly
the SU(3) breaking pattern in the couplings. 
As we will discuss later,
the SU(3) breaking pattern in the couplings
is mainly driven by the quark-mass
terms in the OPE whose determination requires only the zeroth 
moment of the wave function.

The OPE for the correlator in Eq.~(\ref{etwo}) is obtained from Eq.~(\ref{pope})
after replacing $\pi^0 \rightarrow \eta$.  Then the d-quark
component of the  $\eta$ wave function has the opposite sign from 
the pion case, 
as obtained from the soft-meson theorem.  Within the SU(3) limit, 
except for the overall factor of $1/\sqrt{3}$, 
this is the only aspect distinct from the pion sum rule.  This means that
the OPE for $\eta NN$ can be obtained from Eq.~(\ref{popep})
basically by
changing the sign of all the terms except for the last term containing
$m^2_0$. Therefore, the $\eta NN$ sum rule is given by
\begin{eqnarray}
&&g_{\eta N} m_\eta^2 \lambda^2_N e^{-m^2_N/M^2} [ 1+ BM^2]=
\nonumber \\
&&{1\over \sqrt{3}}
\Bigg \{
m_\eta^2 M^4 E_0 (x) \left [
{\langle {\bar q}q \rangle \over 12 \pi^2 f_\eta}
              + {3 f_{3\eta} \over 4\sqrt{2}\pi^2} \right ]
- {2m_q \over f_\eta} \langle {\bar q}q \rangle^2 M^2
- {m_\eta^2 \over 72 f_\eta} \langle {\bar q}q \rangle
\left \langle {\alpha_s \over \pi} {\cal G}^2\right \rangle\nonumber \\
&&+{2m_q\over 3 f_\eta} m_0^2 \langle {\bar q}q \rangle^2 \ \Bigg \}\ .
\label{sum2}
\end{eqnarray}
Here, $B$ denotes the unknown single-pole term, the contribution 
of $N\rightarrow N^*$ plus the one from the derivative of the form factor.
In getting the first and third terms, we have used the
second moment of the $\eta$ wave function 
$\int^1_0 dt~t^2 \varphi_\eta (t) =1/3$
but for the
second and fourth terms containing the quark-mass
we have used the zeroth moment of the $\eta$ wave function. 
To extract $g_{\eta N}$, we have to divide both sides by $m^2_\eta$.
The OPE terms containing $m_q$ will be suppressed by $m_\eta^2$ but
in the other OPE  the $m_\eta^2$ dependence will be canceled.  
As advertised, the SU(3) breaking in $g_{\eta N}$ 
is mainly driven by the quark-mass terms in which the zeroth
moment of the $\eta$ wave function is used. Therefore, our assumption
about the second moment of the $\eta$ wave function is not a
main part in breaking the SU(3) symmetry in the the coupling.

In the SU(3) limit ($f_\eta =f_\pi$, 
$f_{3\eta} = f_{3\pi}$, $m_\eta^2 = m_\pi^2$),
the RHS of Eq.~(\ref{sum2}) becomes,  using the notations introduced 
in Eq.~(\ref{sum1}),
\begin{eqnarray}
{1\over \sqrt{3}}[-{\cal O}_1 - {\cal O}_2 -{\cal O}_3+{\cal O}_4]\ .
\end{eqnarray}
Compared to the OPE for the $\pi NN$ case, 
this reveals interesting aspects of the $\eta NN$ coupling.
Signs of the first three terms are opposite to those in the pion case.
The overall
sign of the $\eta NN$ coupling should be governed by these terms as
the dimension of the operator in last term is the highest in our calculation. 
This indicates that the sign of the $\eta NN$ coupling is 
opposite to the $\pi NN$ coupling !  Furthermore, because of this
sign difference, ${\cal O}_4$ tends to cancel the first two OPE, making
the total OPE strength small.   This cancellation in addition to the overall
suppression factor $1/\sqrt{3}$ leads to the small $\eta NN$ coupling
in the SU(3) limit.

Another but very important aspect is related to the SU(3) relation 
for the $\eta NN$ coupling.
To address this from a simple analysis, postponing a full analysis
to the later sections, 
let us ignore the unknown single-pole terms, $A$ and $B$, in the sum rules 
Eqs.~(\ref{sum1}) and (\ref{sum2}).  
Then the ratio of the two couplings becomes, 
\begin{eqnarray}
{g_{\eta N} \over g_{\pi N}} \sim {1\over \sqrt{3}}~
{-{\cal O}_1 - {\cal O}_2 -{\cal O}_3+{\cal O}_4
\over {\cal O}_1 + {\cal O}_2 +{\cal O}_3+{\cal O}_4}\ .
\label{rel1}
\end{eqnarray}
On the other hand, the $\eta NN$ coupling is known to satisfy the
SU(3) relation~\cite{swart}
\begin{eqnarray}
g_{\eta N} = {g_{\pi N} \over \sqrt{3}} (4 \alpha -1)\ ,
\end{eqnarray}
with $\alpha=F/(F+D)$.  
By comparing our ratio to this relation, we 
immediately see that
\begin{eqnarray}
2\alpha \sim {{\cal O}_4
\over {\cal O}_1 + {\cal O}_2 +{\cal O}_3+{\cal O}_4}\ .
\label{alpha}
\end{eqnarray}
Thus, $\alpha$ is closely related to ${\cal O}_4$, one of
power corrections. 
Because of the neglected unknown
strength, Eq.~(\ref{alpha}) is not an exact 
relation
for $\alpha$.  
Nevertheless, this identification provides an 
important nature of $\alpha$.
We stress that this identification becomes possible because the
sum rules are constructed beyond the chiral limit. 
If the sum rule is constructed using the soft-meson
theorem, the term corresponding to ${\cal O}_4$ does not
participate in the sum rule and we can not make this kind
of identification of $\alpha$. 

Beyond the SU(3) limit, the $\eta NN$ sum rule in Eq.~(\ref{sum2})
has another distinct feature from the $\pi NN$ sum rule.
Even beyond the SU(3) limit, we may still assume that $f_\eta = f_\pi$,
$f_{3\eta} = f_{3\pi}$ as they are not expected
to be changed substantially.   
The most important source for the SU(3) breaking is $m_\eta^2$, which
is much larger than $m_\pi^2$. 
Thus, when both sides of Eq.~(\ref{sum2}) are divided by $m_\eta^2$,
the quark mass terms, ${\cal O}_2$ and ${\cal O}_4$, will be suppressed
by the factor $m_\pi^2/m_\eta^2$ from the corresponding terms in 
Eq.~(\ref{sum1}).  This suppression in addition to the
trivial factor of $1/\sqrt{3}$ will be reflected in the physical
$\eta NN$ coupling.

\section{QCD sum rules for $\pi \Xi\Xi$ and $\eta \Xi\Xi$}
\label{sec:xisum}

The QCD sum rules proposed above have interesting features in the SU(3) 
limit.  
The OPE is basically the same: the OPE for the
$\eta NN$ sum rule
is different from the $\pi NN$ case only by the overall 
factor of $1/\sqrt{3}$
and relative sign of certain terms.  This leads to a simple relation 
for $\alpha$ when the OPE is assumed to be proportional to the coupling.
Thus, the two sum rules in the SU(3) limit can be used
to determine the $F/D$ ratio.
However, for this prediction to be reliable, it is necessary to 
make a consistency
check by calculating other meson-baryon couplings in the SU(3) limit.
For this purpose, we construct QCD sum rules for 
$\pi \Xi\Xi$ and $\eta \Xi\Xi$ in this section.

In constructing the sum rule for $\pi \Xi\Xi$, we use the two-point 
correlation function of the $\Xi$ interpolating field $J_{\Xi}$,
\begin{eqnarray}
\Pi (q,p) = i \int d^4 x e^{i q \cdot x} \langle 0 | T[J_{\Xi} (x) 
{\bar J}_{\Xi} (0)]| \pi^0 (p) \rangle \ ,
\label{pxi}
\end{eqnarray}
with~\cite{qsr} 
\begin{eqnarray}
J_{\Xi} = -\epsilon_{abc} [ s_a^T C \gamma_\mu s_b ] \gamma_5 \gamma^\mu u_c\ .
\end{eqnarray}
Since the $\Xi$ interpolating field is obtained from the nucleon
interpolating field by replacing 
u-quark $\rightarrow$ s-quark and d-quark $\rightarrow$ u-quark, the OPE for
Eq.~(\ref{pxi}) can be obtained directly from Eq.~(\ref{pope}) under the
similar replacements.  Then the term corresponding to 
the last term in Eq.~(\ref{pope}) is zero
because it contains
\begin{eqnarray}
\langle 0| {\bar s}(0) i\gamma_5 s(x) | \pi^0 \rangle =0\ .
\end{eqnarray}
All other OPE now contain the u-quark component instead of the
d-quark component.
According to Eq.~(\ref{twist3}), the u-quark component has 
the opposite sign of the d-quark.  

With these distinctions in mind, we can immediately 
write the sum rule for the $\pi \Xi\Xi$ coupling,
\begin{eqnarray}
&&g_{\pi \Xi}~m_\pi^2 \lambda^2_{\Xi} e^{-m^2_{\Xi}/M^2} [ 1+ CM^2]=
\nonumber \\
&&
m_\pi^2 M^4 E_0 (x) \left [
{\langle {\bar q}q \rangle \over 12 \pi^2 f_\pi}
              + {3 f_{3\pi} \over 4\sqrt{2}\pi^2} \right ]
- {2 m_s\over f_\pi} \langle {\bar s}s \rangle
\langle {\bar q}q \rangle M^2
- {m_\pi^2 \over 72 f_\pi} \langle {\bar q}q \rangle
\left \langle {\alpha_s \over \pi} {\cal G}^2\right \rangle\ .
\label{sum3}
\end{eqnarray}
Again $C$ denotes unknown single pole term representing
the strength $\Xi\rightarrow \Xi^*$.  Here, we identify a
huge SU(3) breaking source in the OPE, $m_s$.
A typical value for $m_s$ is $\sim 150$ MeV, much larger
than $m_u$ or $m_d$.  Another breaking source in the OPE,
the strange quark condensate, is
only about 20 \% smaller than the quark condensate, not 
badly broken from its SU(3) symmetric limit.

To identify $\alpha$ from this sum rule, we take the SU(3) symmetric 
limit again.  The RHS of Eq.~(\ref{sum3}) becomes with the
notations introduced in Eq.~(\ref{sum1}),
\begin{eqnarray}
-{\cal O}_1 -{\cal O}_2 -{\cal O}_3\ .
\end{eqnarray}
Also, we have $m_{\Xi} = m_{N}$.   Another phenomenological parameter 
$\lambda_{\Xi}^2$ also must be equal to the nucleon strength $\lambda_N^2$
in the SU(3) limit.
This parameter  in principle
should be determined from the $\Xi$ mass
sum rule. The $\Xi$ mass sum rule  is different from the nucleon 
mass sum rule only by the terms containing the s-quark mass~\cite{qsr}.  
In the SU(3) 
limit, we have $m_u=m_d=m_s$ and,
\begin{eqnarray}
\lambda_{\Xi}^2 = \lambda_N^2\ .
\end{eqnarray}
Then, as before, by neglecting the unknown single-pole term 
and taking the ratio with the $\pi NN$ sum rule, we obtain
the relation, 
\begin{eqnarray}
{g_{\pi \Xi} \over g_{\pi N}} \sim -{{\cal O}_1 + {\cal O}_2 +{\cal O}_3
\over {\cal O}_1 + {\cal O}_2 +{\cal O}_3+{\cal O}_4}\ .
\end{eqnarray}
Expressing this in terms of $\alpha$ Eq.~(\ref{alpha}) yields
\begin{eqnarray}
{g_{\pi \Xi} \over g_{\pi N}}\sim 2\alpha -1\ .
\label{rel2}
\end{eqnarray}
This exactly matches the SU(3) relation proposed in Ref.~\cite{swart}.
Furthermore, from the OPE structure, we see that
the sign of $g_{\pi \Xi}$ should be opposite to that of $g_{\pi N}$.

Let us turn our discussion onto the $\eta \Xi \Xi$ sum rule.  
For this purpose, we use the correlation function,
\begin{eqnarray}
\Pi (q,p) = i \int d^4 x e^{i q \cdot x} \langle 0 | T[J_{\Xi} (x) 
{\bar J}_{\Xi} (0)]| \eta (p) \rangle \ .
\label{exi}
\end{eqnarray}
In this case,  
a term corresponding to the last term in Eq.~(\ref{pope}) contributes
to the sum rule since the s-quark component with $\eta$ can be written
as 
\begin{eqnarray}
\langle 0| {\bar s}(0) i\gamma_5 s(x) | \eta \rangle 
= +{2\over \sqrt{3}}
{\langle {\bar s} s\rangle \over f_\eta}\int^1_0 dt e^{-itp\cdot x}
\varphi_\eta (t) \ .
\label{seta}
\end{eqnarray}
Again, the coefficient is determined from the soft-meson
theorem while the rest of the $p$-dependence is parametrized in terms
of the twist-3 wave function. Only the zeroth moment of this
wave function, which is purely governed by the soft-meson theorem,
contributes to this sum rule and therefore we don't need to make
a further assumption regarding the second moment of this wave function.
All other OPE contain the u-quark component with $\eta$.

By noting the similarities and distinctions from the $\pi \Xi \Xi$ sum rule,
it is straightforward to write down the sum rule for $\eta \Xi \Xi$,
\begin{eqnarray}
&&g_{\eta \Xi}~ m_\eta^2 \lambda^2_\Xi e^{-m^2_\Xi/M^2} [ 1+ DM^2]=
\nonumber \\
&&{1\over \sqrt{3}}
\Bigg \{
m_\eta^2 M^4 E_0 (x) \left [
{\langle {\bar q}q \rangle \over 12 \pi^2 f_\eta}
              + {3 f_{3\eta} \over 4\sqrt{2}\pi^2} \right ]
- {2m_s \over f_\eta} \langle {\bar s}s \rangle \langle {\bar q}q \rangle M^2
- {m_\eta^2 \over 72 f_\eta} \langle {\bar q}q \rangle
\left \langle {\alpha_s \over \pi} {\cal G}^2\right \rangle\nonumber \\
&&-{2 m_0^2 \over 3 f_\eta} \left [ m_s \langle {\bar q}q \rangle
\langle {\bar s}s \rangle +
m_q \langle {\bar s}s \rangle^2 \right ] \ \Bigg \}\ .
\label{sum4}
\end{eqnarray}
Again, we identify the SU(3) breaking sources, $m_s$,
$\langle {\bar s} s\rangle$, $f_\eta$ and $f_{3\eta}$.
The most important source for the SU(3) breaking is $m_s$.  The
terms containing $m_s$ are obtained by using the zeroth moment of the $\eta$
wave function whose value is well-fixed by the soft-meson theorem.

Using the notations introduced in Eq.~(\ref{sum1}), the RHS of 
Eq.~(\ref{sum4}) in {\it the SU(3) limit}
becomes
\begin{eqnarray}
{1\over \sqrt{3}}[-{\cal O}_1 - {\cal O}_2 -{\cal O}_3-2{\cal O}_4]\ .
\end{eqnarray}
This OPE indicates that $g_{\eta \Xi}$ has the opposite sign of $g_{\pi N}$.
By neglecting the unknown constant $D$ within the SU(3) limit and taking 
the ratio with the $\pi NN$ sum rule, we obtain
\begin{eqnarray}
{g_{\eta \Xi} \over g_{\pi N}} \sim -{1\over \sqrt{3}} 
{ {\cal O}_1 + {\cal O}_2 +{\cal O}_3+2{\cal O}_4 \over 
{\cal O}_1 + {\cal O}_2 +{\cal O}_3+{\cal O}_4}\ .
\label{rel3}
\end{eqnarray}
In terms of $\alpha$ Eq.~(\ref{alpha}), 
the ratio becomes 
\begin{eqnarray}
 {g_{\eta \Xi} \over g_{\pi N}}  \sim -{1\over \sqrt{3}} (1+2\alpha)
\end{eqnarray}
matching the SU(3) relation of Ref.~\cite{swart} exactly again.  
Therefore, our identification of $\alpha$ as given in Eq.~(\ref{alpha})
suggests that our OPE for the $\pi \Xi \Xi$ and $\eta \Xi \Xi$
couplings are consistent with 
the SU(3) relations.

\section{QCD sum rules for $\pi \Sigma\Sigma$ and $\eta \Sigma\Sigma$}
\label{sec:sigmasum}

Another examples where our formalism is directly applicable
are the $\pi \Sigma\Sigma$ and $\eta \Sigma\Sigma$
couplings.  To calculate the couplings, we simply substitute
the nucleon interpolating field with the $\Sigma$ interpolating
field, $J_p \rightarrow J_\Sigma$.  
For the $\pi \Sigma \Sigma$ sum rule,  we need to consider
\begin{eqnarray}
i \int d^4 x e^{i q \cdot x} \langle 0 | T[J_\Sigma (x) 
{\bar J}_\Sigma (0)]| \pi^0 (p) \rangle \ ,
\label{stwo}
\end{eqnarray}
with~\cite{qsr}
\begin{eqnarray}
J_{\Sigma} = 
\epsilon_{abc} [ u_a^T C \gamma_\mu u_b ] \gamma_5 \gamma^\mu s_c\ .
\end{eqnarray}
This interpolating field can be obtained from the nucleon
interpolating field by replacing,
\begin{eqnarray}
{\rm d-quark} \rightarrow {\rm s-quark}\ .
\end{eqnarray}
This means that the OPE in this case can be obtained from
Eq.~(\ref{pope}) by the same replacement and, as the s-quark component
with a pion is zero, only the term corresponding to the last term in
Eq.~(\ref{pope}) will give a nonzero contribution.
Thus, the sum rule for $\pi \Sigma \Sigma$ is
\begin{eqnarray}
g_{\pi \Sigma}~m_\pi^2 \lambda^2_\Sigma e^{-m^2_\Sigma/M^2} [ 1+ EM^2]=
{m_0^2\over 3 f_\pi } \left [ m_q \langle {\bar s}s \rangle 
\langle {\bar q}q \rangle + m_s \langle {\bar q}q \rangle^2 \right ]\ .
\label{sum5}
\end{eqnarray}
As $m_s$ changes substantially in the SU(3) breaking limit, 
the OPE undergoes a substantial change as we go from the
SU(3) symmetric limit to its breaking limit.
In the SU(3) symmetric limit [$m_s=m_q$, $\langle {\bar s} s \rangle
=\langle {\bar q} q\rangle$],  the
RHS, which is then equal to ${\cal O}_4$, satisfies the SU(3) relation
if Eq.~(\ref{alpha}) is used, that is,
\begin{eqnarray}
{g_{\pi \Sigma} \over g_{\pi N}}  \sim 2 \alpha\ .
\label{rel4}
\end{eqnarray}
Note, the OPE is positive, $\alpha >0$. 

Now, for the $\eta \Sigma \Sigma$ case, we can similarly
proceed using the correlator
\begin{eqnarray}
i \int d^4 x e^{i q \cdot x} \langle 0 | T[J_\Sigma (x) 
{\bar J}_\Sigma (0)]| \eta (p) \rangle \ .
\end{eqnarray}
In this case, similarly
as the $\eta \Xi \Xi$ case, the s-quark component have
a nonzero value with $\eta$ [see Eq.(\ref{seta}).].  
Straightforward calculations yield,
\begin{eqnarray}
&&g_{\eta \Sigma}~m_\eta^2 \lambda^2_\Sigma e^{-m^2_\Sigma /M^2} [ 1+ FM^2]=
\nonumber \\
&&{1\over \sqrt{3}}
\Bigg \{
-2 m_\eta^2 M^4 E_0 (x) \left [
{\langle {\bar s}s \rangle \over 12 \pi^2 f_\eta}
              + {3 f_{3\eta} \over 4\sqrt{2}\pi^2} \right ]
+ {4m_q \over f_\eta} \langle {\bar s}s \rangle \langle {\bar q}q \rangle M^2
+ {m_\eta^2 \over 36 f_\eta} \langle {\bar s}s \rangle
\left \langle {\alpha_s \over \pi} {\cal G}^2\right \rangle\nonumber \\
&&+{m_0^2\over 3 f_\eta}\left [m_q \langle {\bar s}s \rangle 
\langle {\bar q}q \rangle + m_s \langle {\bar q}q \rangle^2 \right ] 
\Bigg \}\ .
\label{sum6}
\end{eqnarray}
In getting  the first and third terms in the OPE, we have
used the second moment of the $\eta$ wave function, 
$1/3$.  This assumption is made in analogy with the
pion wave function.  However, since we are dealing with the
s-quark component with $\eta$ in this case,  our assumption for the
$\eta$ wave function for the second moment has been further extended.
Once again in the SU(3) limit, the RHS takes the form
\begin{eqnarray}
{1 \over \sqrt{3}} \left [2{\cal O}_1+2{\cal O}_2+2{\cal O}_3+
{\cal O}_4\right ]\ ,
\end{eqnarray}
which, when combined with Eq.~(\ref{alpha}),
yields the SU(3) relation for $\eta \Sigma \Sigma$,
\begin{eqnarray}
{g_{\eta \Sigma} \over g_{\pi N}} \sim
{2\over \sqrt {3}} (1-\alpha)\ ,
\label{rel5}
\end{eqnarray}
matching again the SU(3) relation of de Swart~\cite{swart}.

\section{Analysis in the SU(3) limit $-$ determination of the $F/D$ ratio}
\label{sec:anal}

So far, we have presented the sum rules for $\pi NN$, $\eta NN$
$\pi \Xi \Xi$, $\eta \Xi \Xi$, $\pi \Sigma \Sigma$ 
and $\eta \Sigma\Sigma$
beyond the chiral limit. As far as the OPE is concerned,
the sum rules satisfy the SU(3) relations, indicating
that they are not independent.  They are related through
the SU(3) rotations.  
One assumption made to the second moment of the $\eta$ wave
function, which is taken to be the same as the one from the
pion wave function $\int^1_0 dt~t^2 \varphi_\eta (t) =1/3$,
is valid in this consideration, perhaps motivating its use even beyond
the SU(3) symmetric limit.  From the consistency with the SU(3) relations, we
have identified the OPE responsible for the $F/D$ ratio. In reaching
this identification, it is important to construct the QCD sum
rules beyond the chiral limit.

\begin{figure}
\caption{ The Borel mass dependence of $a+bM^2$ for $\pi NN$
and $\eta NN$ sum rules in the SU(3) symmetric case.
The continuum threshold $S_0 = 2.07 $ GeV$^2$, corresponding
to the Roper resonance, is used for the solid lines.
To see the sensitivity to the continuum threshold,
the dashed-lines with $S_0 =2.57 $ GeV$^2$ are also
plotted.
In the case of $\pi NN$, the continuum gives 2\%
corrections at $M^2=1$ GeV$^2$. 
}
\label{fig1}

\setlength{\textwidth}{6.1in}   
\setlength{\textheight}{9.in}  
\centerline{%
\vbox to 2.4in{\vss
   \hbox to 3.3in{\includegraphics{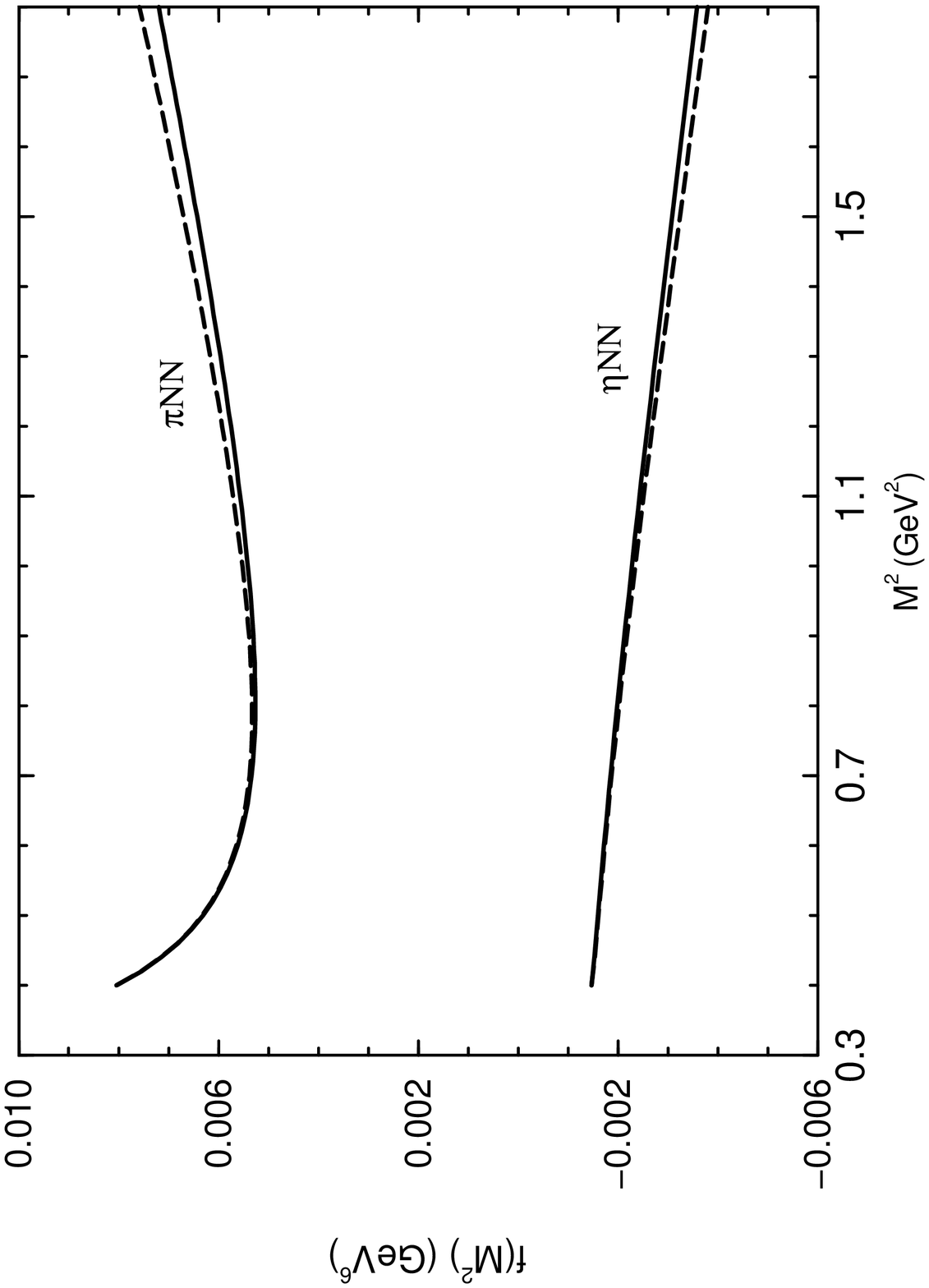}\hss}}
}
\vspace{100pt}
\end{figure}

In this section, we determine the $F/D$ ratio from 
the sum rules Eqs. (\ref{sum1}) (\ref{sum2}) (\ref{sum3}) (\ref{sum4}) 
(\ref{sum5}) (\ref{sum6}).  As the $F/D$ ratio is a parameter
defined in the SU(3) limit, we consider the sum rules
in {\it the SU(3) limit}.  
In this limit, we have exact relations,
\begin{eqnarray}
\lambda_N^2 = \lambda_\Xi^2 = \lambda_\Sigma^2\ ,
\label{sucoup}
\end{eqnarray}
as all the baryonic
mass sum rules are equal~\cite{qsr}.
In our analysis, we use the standard QCD parameters,
\begin{eqnarray}
\langle {\bar q} q \rangle = -(0.23~ {\rm GeV})^3\;; \quad
\left \langle {\alpha_s \over \pi} {\cal G}^2\right \rangle
= (0.33~{\rm GeV})^4\;; \quad m_0^2=0.8~{\rm GeV}^2\ .
\end{eqnarray}
We arrange each sum rule into the form
\begin{eqnarray}
a+ b M^2 = f(M^2)
\end{eqnarray}
by dividing both sides of each sum rule by the meson mass squared and the
exponential factor in the phenomenological side.
$b M^2$ indicates the contributions from
$N \rightarrow N^*$ (or $\Xi \rightarrow \Xi^*$, 
$\Sigma \rightarrow \Sigma^*$).  
Specifically, in the $\pi NN$ sum rule Eq.~(\ref{sum1}),
we divide both sides by $m_\pi^2~e^{-m^2_N/M^2}$.
Thus, in this sum rule, $a = g_{\pi N} \lambda_N^2$, 
$b=g_{\pi N} \lambda_N^2 A$ and the RHS becomes
\begin{eqnarray}
f(M^2)={ {\cal O}_1+{\cal O}_2+{\cal O}_3+{\cal O}_4 \over
m_\pi^2~e^{-m^2_N/M^2}}\ .
\label{asum1}
\end{eqnarray}
Recall that ${\cal O}_1$ and ${\cal O}_3$ contain $m_\pi^2$. 
The quark-mass $m_q$  in ${\cal O}_2$ and ${\cal O}_4$,
with the use of the 
Gell-Mann$-$Oakes$-$Renner relation, can be converted
to $m_\pi^2$, which is then canceled with
another $m_\pi^2$  in the denominator. Therefore, the sum rule in 
its final form
does not depend on the quark-mass or $m_\pi^2$.
The parameters $a$ and $b$ will be determined by fitting $f(M^2)$ 
with a straight line within a Borel window.  
Similarly constructed Borel curves for all sum rules 
$f(M^2)$  are shown in Fig.~\ref{fig1},
Fig.~\ref{fig2} and Fig.~\ref{fig3}.
A common feature is that the continuum threshold does not affect 
the Borel curve much.

In the fitting process, we need to choose an appropriate
Borel window.  As can be seen from the figures, depending
on the Borel window we choose, we would get 
different values for the parameters $a$ and $b$.
But as long as the Borel window is chosen 
for $M_{min}^2 \ge 0.7$ GeV$^2$,  all Borel curves
are relatively well-fitted by straight lines, reducing the
sensitivity to the Borel window. Nonetheless, 
a more important claim about the Borel windows can
be made from the OPE structure of each sum rule.
As we have demonstrated in earlier sections, 
the OPE of each sum rule in the SU(3) limit satisfies the SU(3) relation.
This means that as far as the OPE is concerned, all sum rules
are related by the SU(3) rotations: they are not independent.
A point in a Borel curve at a specific Borel mass
is transformed under the 
SU(3) rotation to a point of an other Borel curve defined
at the same Borel mass.  
It implies that once a Borel window is fixed in one sum rule,
the same Borel window must be applied to the other sum rules as
they are not independent.
This claim can be justified if the extracted $F/D$ ratio is
independent of the sum rule considered.

We fix the common Borel window from the $\pi NN$ sum rule.
According to the analysis in Ref.~\cite{hung3},
the Borel window $0.65 \le M^2 \le 1.24$ GeV$^2$ is obtained as
a common window for the $\pi NN$ sum rule and the chiral-odd nucleon mass
sum rule.  It provides
$g_{\pi N}$ close to its empirical value. 
Using this Borel window to other sum rules, we determine the 
parameters $a$ and $b$ from  each Borel curve. 
The parameter $a$ is obtained from the intersection of the
best-fitting curve with the y-axis. The slope of the best fitting 
curve yields the parameter $b$. They are listed in table~\ref{tab1}.
From the first column, we see that the signs of the couplings
obtained from the best-fitting method are consistent with
our naive analysis given in Eqs.~(\ref{rel1}), (\ref{rel2}),
(\ref{rel3}), (\ref{rel4}), (\ref{rel5}). Specifically,
\begin{eqnarray}
&&g^{(S)}_{\eta N} < 0\;; \quad g^{(S)}_{\pi \Xi} < 0
\;; \quad g^{(S)}_{\eta \Xi} <0 \nonumber \\
&& g^{(S)}_{\pi \Sigma} > 0\;; \quad g^{(S)}_{\eta \Sigma} > 0 \ .
\label{sign}
\end{eqnarray}
Here the superscript $(S)$ indicates the couplings in the SU(3) limit.
These signs restrict the range of $\alpha [\equiv F/(F+D)]$,
\begin{eqnarray}
0 < \alpha < {1\over 4} \rightarrow 0 < {F\over D} < {1\over 3}\ .
\label{constraint}
\end{eqnarray}
This is a constraint for the $F/D$ ratio to be satisfied within 
our QCD sum rule analysis.

From the ratios provided in the fourth column, we determine
$\alpha$ 
from each sum rule since we know the SU(3)
relation for each coupling~\cite{swart}\footnote{ Note, according
to Eq.~(\ref{sucoup}), all strengths of the interpolating fields to
the baryons are  the same. Therefore,  we can think of the parameter
$a$ as the coupling times the multiplicative factor common
to all sum rules.}.
The five ratios
presented in table~\ref{tab1} consistently
give $\alpha=0.175$. This  justifies  the use of the common Borel
window for all sum rules.  The $F/D$ ratio from this value is
$0.212$.
To see the sensitivity to the Borel window, we blindly
shift the common Borel window to $0.9 \le M^2 \le 1.5$ GeV$^2$.
Of course, in this window, we would not obtain the $\pi NN$ coupling
consistent with its empirical value. Nevertheless,
the $F/D$ ratio from this Borel window is 0.196, which is
not far from the one above.  From this analysis, we can safely
claim that 
\begin{eqnarray}
F/D\sim 0.2\ .
\end{eqnarray}
This is about a factor of 3 smaller than
what the SU(6) predicts, $F/D = 2/3$ or the more recent 
value~\cite{rat}, $F/D\sim 0.57$.
It is interesting to see that
these values from other studies even badly violate the
constraint from our study Eq.~(\ref{constraint}). 
The small $F/D$ ratio makes sense in our approach
because it is generated by the highest dimensional operator in
the OPE.

From the ratios in table~\ref{tab1},
we can also calculate the couplings in the SU(3) limit if the
$\pi NN$ coupling is given.
Using $g_{\pi N} = 13.4$, we obtain from the fourth column in
table~\ref{tab1},
\begin{eqnarray}
&&g^{(S)}_{\eta N} = -2.3 \;; \quad g^{(S)}_{\pi \Xi}=-8.7 \;; 
\quad g^{(S)}_{\eta \Xi} =-10.5
\nonumber\ \\
&&g^{(S)}_{\pi \Sigma} =4.7 \;; \quad g^{(S)}_{\eta \Sigma} =12.8\ .
\label{symcoup}
\end{eqnarray}
Note, the signs of the couplings are relative to $g_{\pi N}$.
Each coupling, when it is combined with the first column of 
table~\ref{tab1}, consistently
yields the strength,
\begin{eqnarray}
\lambda_N^2\sim 3.5 \times 10^{-4}~{\rm GeV}^6\ ,
\end{eqnarray}
which of course should be equal to $\lambda_\Xi^2$ and
$\lambda_\Sigma^2$ in the SU(3) limit.  This
is the first determination of $\lambda_N$ without explicit use
of the nucleon mass sum rules.  For later discussions, we 
close this section by listing the ratios of 
$\eta$-baryon couplings over pion-baryon couplings obtained in the
SU(3) limit,
\begin{eqnarray}
{g^{(S)}_{\eta N} \over g_{\pi N}} =-0.17\;; \quad
{g^{(S)}_{\eta \Xi} \over g^{(S)}_{\pi \Xi}} =1.2\;; \quad
{g^{(S)}_{\eta \Sigma} \over g^{(S)}_{\pi \Sigma}} = 2.71\ .
\label{srat}
\end{eqnarray}
Note, we have not put the superscript on $g_{\pi N}$
because it is independent of whether or not the SU(3) limit is taken.
The magnitude of $g^{(S)}_{\eta N}$ is a lot smaller than $g^{(S)}_{\pi N}$,
while $g^{(S)}_{\eta \Xi}$ and $g^{(S)}_{\eta \Sigma}$ are larger than
the corresponding pion-baryon couplings.

\section{Meson-baryon couplings$-$qualitative analysis}
\label{sec:anal2}

Having established the $F/D$ ratio in the SU(3) limit, we now
move on to an analysis {\it beyond} the SU(3) limit.
Due to the symmetry breaking, we 
have different baryonic masses, meson masses,
and $\lambda_N \ne \lambda_\Xi \ne \lambda_\Sigma$.
Moreover, some QCD parameters change from their values in the symmetric
limit. 
This means that all sum rules are not simply related by
the SU(3) rotations and a separate Borel analysis is necessary 
in each sum rule.
However, in predicting the couplings,
we have several limitations.  First, as mentioned,
the assumption used for the second moment of the $\eta$ wave function
may not be valid in the breaking limit. 
We speculate from
the soft-meson limit that corrections to this assumption are small but
the soft-meson theorem may not be strictly valid as mesons
become heavier.
In addition, we do not have clear restrictions on the QCD parameters
$f_\eta$, $f_{3\eta}$, $m_s$ and
$\langle {\bar s} s \rangle$.  The $\eta$ decay constant $f_\eta$ is
known to be about 20 \% larger from its SU(3) value but
$f_{3\eta}$ is not well under control. 
The standard values for $m_s$ and $\langle {\bar s} s \rangle$ are
\begin{eqnarray}
m_s = 150~{\rm MeV}\;; \quad 
\langle {\bar s} s \rangle = 0.8 \langle {\bar q} q \rangle\ .
\end{eqnarray}
Even though we will use these values in this work,
there are still some on-going discussions on these values~\cite{malt2,antonio}.
Moreover, there is an additional source of the SU(3) breaking
driven by the $\eta - \eta^\prime$ mixing.  Some studies~\cite{nemoto}
suggest that $\eta$ is almost $\eta_8$ but there is no
consensus on this point. 
Therefore, our analysis in this section should be regards as ``qualitative''.

We will take $f_\pi = f_\eta$, $f_{3\pi}=f_{3\eta}$, 
and ignore the mixing between $\eta - \eta^\prime$.
The most important source for the SU(3) breaking in the QCD side is
the strange quark mass $m_s$.  Compared to the up-(or down-)quark
mass, $m_s$ is very large, more than 20 times of $m_q$. 
In the hadronic side, baryonic masses, meson masses, and strength
of the interpolating fields to the physical baryons will
be changed in the breaking.  In this section, we will study
how these breakings are reflected in the couplings.

Figure~\ref{fig4} shows the Borel curve for the $\eta NN$
coupling. It is almost a straight line with respect
to $M^2$. Recall that this curve is obtained by dividing
Eq.~(\ref{sum2}) with $m_\eta^2 = (547~{\rm MeV})^2$ in addition to the
exponential factor of $e^{-m_N^2/M^2}$.  The quark-mass
terms in the OPE will be suppressed by the factor $m_\pi^2/m_\eta^2$
compared to the corresponding terms in the SU(3) symmetric case.  
This will shift the Borel curve upward as shown in 
Fig.~\ref{fig4}, which gives
small value $a\sim-0.00022$ in table~\ref{tab2}.
Since $a$ is very small, the relative sensitivity to the 
continuum threshold or the Borel window becomes large in this case.
Nevertheless, by dividing it with the $a$ from the $\pi NN$ sum rule and
using the empirical value $g_{\pi N}=13.4$, we obtain
$g_{\eta N}=-0.63$, negative value but its magnitude practically 
consistent with zero.  

A remarkable breaking effect can be observed in the $\pi \Xi\Xi$
Eq.~(\ref{sum3}).  As $m_s$ involved in the OPE is 
substantially increased in the SU(3) breaking limit,
the quark-mass term will be enhanced by a factor of 20 while
other OPE terms remain the same. 
A further enhancement of the Borel curve comes
from the exponential factor $e^{m_\Xi^2/M^2}$.
The resulting Borel curve is shown in Fig.~\ref{fig5}.
Around the resonance mass, $m_\Xi^2\sim 1.73$ GeV$^2$,
the Borel curve is almost flat, indicating that the unknown
single pole term is small.  This kind of strong enhancement
of the Borel curve is not observed in the
$\eta \Xi \Xi$ case.  In this case, the dividing factor $m_\eta^2$,
much larger than $m_\pi^2$, compensates the strong
enhancement coming from $m_s$. 
Similar behaviors can be observed in Fig.~\ref{fig6}
for $\pi \Sigma \Sigma$
and $\eta \Sigma \Sigma$. Here the enhancement of the $\pi \Sigma \Sigma$ 
Borel curve due to $m_s$
is not so strong as $\pi \Xi \Xi$ because the OPE contains
only one term.   

The best fitting parameters for $a$ and $b$ from each sum rule
are listed in table~\ref{tab2}.
Recall that the numbers in the first column is the couplings 
multiplied by the strengths of the
interpolating fields to the baryons, $\lambda_N^2$ 
(or $\lambda_\Xi^2$ or $\lambda_\Sigma^2$ depending on the sum rule).
Since the SU(3) symmetry is broken, the strengths are not
equal. To see the SU(3) breaking effects, we compare the
first column in table~\ref{tab1} with the one in table~\ref{tab2} and 
obtain the ratios
\begin{eqnarray}
&&{g^{(B)}_{\eta N} \over g^{(S)}_{\eta N}} =0.275\;; \quad
{g^{(B)}_{\pi \Xi} [\lambda^{(B)}_\Xi]^2 
\over g^{(S)}_{\pi \Xi}\lambda_N^2} =33.6\;; \quad
{g^{(B)}_{\eta \Xi} [\lambda^{(B)}_\Xi]^2 
\over g^{(S)}_{\eta \Xi}\lambda_N^2} =1.63\ ,\nonumber \\
&&
{g^{(B)}_{\pi \Sigma} [\lambda^{(B)}_\Sigma]^2 
\over g^{(S)}_{\pi \Sigma}\lambda_N^2} =12.34\;; \quad
{g^{(B)}_{\eta \Sigma} [\lambda^{(B)}_\Sigma]^2 
\over g^{(S)}_{\eta \Sigma}\lambda_N^2} =0.447\ .
\label{sbrat}
\end{eqnarray}
The superscript $(B)$ denotes that the couplings in the
SU(3) breaking limit.  
In the denominators, we have $\lambda_N^2$ instead
of the strength for the corresponding interpolating field because
all strengths are the same in the SU(3) limit.
In the ratio for the $\eta NN$ coupling, the strength
of the nucleon interpolating field $\lambda_N^2$ has been
canceled as it is blind to the SU(3) symmetry.
Most ratios show the huge SU(3) breaking. This
is mainly driven by $m_s$ in the QCD side and the meson masses
in the phenomenological side. 
To see the breaking reflected only in the couplings, we need to
eliminate the strength of each baryon to its interpolating
field. 
A recent study~\cite{dey} suggests that the strength $\lambda_B^2$ for
the baryon $B$ with mass $M_B$  scales like $\lambda_B^2 \sim
C M_B^6$ with some constant $C$.  Of course this scaling
needs to be confirmed by further studies but nevertheless  
using this information in our sum rules,
we obtain  the following ratios,
\begin{eqnarray}
&&{g^{(B)}_{\eta N} \over g^{(S)}_{\eta N}} =0.275\;; \quad
{g^{(B)}_{\pi \Xi}  
\over g^{(S)}_{\pi \Xi}} =4.45\;; \quad
{g^{(B)}_{\eta \Xi}  
\over g^{(S)}_{\eta \Xi}} =0.22\ ,\nonumber \\
&&
{g^{(B)}_{\pi \Sigma}  
\over g^{(S)}_{\pi \Sigma}} =2.99\;; \quad
{g^{(B)}_{\eta \Sigma}  
\over g^{(S)}_{\eta \Sigma}} =0.11\ .
\label{sbrat2}
\end{eqnarray}
These clearly show huge SU(3) breaking in the couplings.
Ref.~\cite{chiu} provides different values for 
the strengths $\lambda_\Sigma$ and $\lambda_\Xi$, which do not seem
to satisfy the scaling~\cite{dey}.  These strengths from Ref.~\cite{chiu} 
if used in our work provide larger values for the 
ratios in Eq.~(\ref{sbrat2}).  In this case, the SU(3) breaking
in the couplings becomes more drastic. Therefore, the scaling
$\lambda_B^2 \sim C M^6_B$ provides a mild SU(3) breaking in the
couplings even though it is still large.  
By combining Eq.~(\ref{sbrat2}) with the couplings in the SU(3) limit
Eq.~(\ref{symcoup}), we calculate the couplings
beyond the SU(3) limit  and present them in table~\ref{tab3}
as well as the ones in the SU(3) limit.  The suppression
of $g_{\eta NN}$
seems to support the results of Ref.~\cite{weber} and is
severe than the one from Bonn potential~\cite{bonn}.  The couplings
with hyperons do not agree with Nijmegen potential model~\cite{rijken}.
Nijmegen potential is based on SU(3) symmetry and the
SU(3) breaking enters in the model perturbatively. The couplings
determined by fitting the experimental $YN$ scatterings
seem to be consistent with SU(3) symmetry~\cite{rijken}.
However, the hyperon-interaction data are rather scarce to determine
the couplings reliably and it will be interesting in future to
see how our findings affect the analysis of Nijmegen potential.
To go that direction however we need to make a further study of 
our approach and  solidify our results.
As we have mentioned, our
results beyond the SU(3) limit at this stage should be regarded as
qualitative due to certain assumptions made in the analysis.

To see the SU(3) breaking without
explicit use of the informations from Ref.~\cite{dey},
we also calculate the ratios,
\begin{eqnarray}
{g^{(B)}_{\eta N} \over g_{\pi N}} =-0.05\;; \quad
{g^{(B)}_{\eta \Xi} \over g^{(B)}_{\pi \Xi}} =0.06\;; \quad
{g^{(B)}_{\eta \Sigma} \over g^{(B)}_{\pi \Sigma}} = 0.1\ .
\label{brat}
\end{eqnarray}
When these ratios are compared with the corresponding
ones in the SU(3) limit Eq.~(\ref{srat}), 
we again notice that
the SU(3) breaking effects are huge in the couplings.
The three ratios are consistently smaller in magnitude than their 
corresponding values
in the SU(3) symmetric limit.  The suppression 
is severe in the ratios, $g_{\eta \Xi} / g_{\pi \Xi}$ and
$g_{\eta \Sigma} / g_{\pi \Sigma}$. 
The pion-baryon couplings are enhanced by the strange-quark
mass in the OPE, while in the $\eta$-baryon couplings, this
enhancement is reduced by the overall dividing factor $m_\eta^2$.
When the ambiguity in the form factor is considered, the 
$\eta$-baryon couplings
can be increased by 13 \%.  More softer cut-off like $\Lambda \sim
1$ GeV increases the coupling slightly more. But even so, it
does not change our conclusion that the SU(3) breaking is huge.
 Further changes are expected from
the uncertainties in $m_s$,  $\eta-\eta^\prime$ mixing and so forth.
Nevertheless, the trend that we have observed, especially the claim 
that there is huge SU(3) breaking in the the ratios Eqs.~(\ref{srat})
(\ref{brat}), is
expected to be maintained even if we take into account 
the limitations of our approaches.

Another important finding in our work is the relative signs of the
couplings with respect to $g_{\pi N} > 0$.  The signs provided in
Eq.~(\ref{sign}) are preserved even in the SU(3) breaking limit. 
\begin{eqnarray}
&&g^{(B)}_{\eta N} < 0\;; \quad g^{(B)}_{\pi \Xi} < 0
\;; \quad g^{(B)}_{\eta \Xi} <0 \nonumber \\
&& g^{(B)}_{\pi \Sigma} > 0\;; \quad g^{(B)}_{\eta \Sigma} > 0\ .
\end{eqnarray}
 For the $\eta NN$ coupling, the sign should be opposite of $g_{\pi N}$, because
the highest dimensional operator should be smaller in
magnitude than the leading OPE terms.  For the other
couplings, the signs can be simply
read off from each OPE. Certainly, in the SU(3) limit,
these signs are crucial in explaining the consistency of each OPE with
the SU(3) relation for the corresponding coupling.

\section{Summary}
\label{sec:sum}

In this work, we have developed QCD sum rules beyond
the chiral limit for the diagonal meson-baryon couplings,
$\pi NN$, $\eta NN$, $\pi \Xi \Xi$, $\eta \Xi \Xi$,
$\pi \Sigma \Sigma$ and $\eta \Sigma \Sigma$.
We have assumed the second moment of the twist-3 $\eta$ wave
function to be the same as the one from the pion wave function.
This should be exact in the SU(3) limit. 
The most important finding in this work is that 
the OPE structures match the SU(3) relations
for the couplings in the SU(3) limit.
Going beyond the chiral limit is crucial for this
identification.  Thus, we have identified the
OPE responsible for the $F/D$ ratio.  From a Borel
analysis, it was found to be around $F/D \sim 0.2$
strongly disagreeing with the SU(6) prediction.  In future, it will be
useful to do similar calculations for other Dirac structures~\cite{hung2,hung1}
and see if consistent results can be obtained.
In the SU(3) breaking case,  we have calculated the ratios,
the couplings divided by
the corresponding values in the SU(3) limit.  
The ratios, even though the scaling law is used for 
the baryon strength $\lambda_B^2\sim M^6_B$, suggest that the 
couplings violate  the SU(3) symmetry strongly.  
Also we have presented the 
ratios of $\eta$-baryon couplings to pion-baryon couplings, 
which does not require an assumption for the strength $\lambda_B^2$.
We found that $\eta$-baryon couplings are much
smaller than pion-baryon couplings. Compared to
the corresponding ratio in the SU(3) limit, we have
found that huge SU(3) breaking exists in the couplings.
In future, it will be interesting to explore this aspect in hyperon-nucleon
interactions.

\acknowledgments
This work is supported in part by the 
Grant-in-Aid for JSPS fellow, and
the Grant-in-Aid for scientific
research (C) (2) 11640261 
of  the Ministry of Education, Science, Sports and Culture of Japan.
The work of  H. Kim is also supported by Research Fellowships of
the Japan Society for the Promotion of Science.
The work of S. H. Lee is supported by  the 
Korean Ministry of Education through grant no. 98-015-D00061.

\begin{table}
\caption{
The best-fitting values for the parameters $a$ and $b$ in 
the SU(3) symmetric limit are listed for
each sum rule within the
Borel window $0.65 \le M^2 \le 1.24$ GeV$^2$.   
In the fourth column, we present ratios of each
coupling divided by $g_{\pi N}$, which
are directly related to the $F/D$ ratio.
The obtained value for the $F/D$ ratio is $0.212$.}

\begin{center}
\begin{tabular}{cccc}
& $a$ (GeV$^6$) & $b$ (GeV$^4$) & coupling$/g_{\pi N}$ \\
\hline\hline
$\pi NN$ & $0.00464$ & $0.00084$ & $ 1$ \\
$\eta NN$ & $-0.0008$ & $-0.00146$ & $-0.172 $  \\
$\pi \Xi\Xi$ & $-0.00302$ & $-0.00171$ & $-0.651$  \\
$\eta \Xi\Xi$ & $-0.00362$ & $-0.00002$ & $-0.78$  \\
$\pi \Sigma\Sigma$ & $0.00163$ & $-0.00087$ & $0.351$  \\
$\eta \Sigma\Sigma$ & $0.00442$ & $0.00147$ & $0.953$  
 \\
\end{tabular}
\end{center}
\label{tab1}

\end{table}

\begin{table}
\caption{The best-fitting values for the parameters $a$ and $b$ beyond
the SU(3) symmetric limit are listed for
each sum rule within the
Borel windows taken around the resonance masses.
For $\pi \Sigma \Sigma$ case, there is no continuum
contribution because the OPE does not contain the
perturbative part.}

\begin{center}
\begin{tabular}{ccccc}
& $a$ (GeV$^6$) & $b$ (GeV$^4$) & Borel window (GeV$^2$)& $S_0$ (GeV$^2$) \\
\hline\hline
$\eta NN$ & $-0.00022$ & $-0.00097$ & $0.65 - 1.24$ & $2.07$ \\
$\pi \Xi\Xi$ & $-0.1015$ & $-0.00202$ & $1.53 - 1.93$ & $3.$  \\
$\eta \Xi\Xi$ & $-0.0059$ & $-0.001$ & $1.53 - 1.93$ & $3.$  \\
$\pi \Sigma\Sigma$ & $0.02012$ & $-0.0072$ & $1.21 - 1.61$ & -  \\
$ \eta \Sigma \Sigma$ & $0.00202$ & $0.00178$ & $1.21 - 1.61$ & $3.$ 
 \\
\end{tabular}
\end{center}
\label{tab2}

\end{table}

\begin{table}
\caption{ Meson-baryon diagonal couplings in the SU(3) limit
and beyond the SU(3) limit are presented. As we have discussed
in the text, the values beyond the SU(3) limit should be regarded
as qualitative. The $\pi NN$ coupling in the first line is 
the empirical value. }

\begin{center}
\begin{tabular}{ccc}
& SU(3) limit  & Beyond the SU(3) limit \\
\hline\hline
$g_{\pi N}$ & $13.4$ & $13.4$ \\
$g_{\eta N}$ & $-2.3$ & $-0.63$ \\
$g_{\pi \Xi}$ & $-8.7$ & $-38.7$ \\
$g_{\eta \Xi}$ & $-10.5$ & $-2.3$ \\
$g_{\pi \Sigma}$ & $4.7$ & $14.1$  \\
$ g_{\eta \Sigma} $ & $12.8$ & $1.4 $ 
 \\
\end{tabular}
\end{center}
\label{tab3}

\end{table}

\begin{figure}
\caption{ The Borel curves for $\pi \Xi \Xi$ and
$\eta \Xi \Xi$ in the SU(3) limit. The dashed lines show the
sensitivity to the continuum threshold.}
\label{fig2}
\end{figure}

\setlength{\textwidth}{6.1in}   
\setlength{\textheight}{9.in}  
\begin{figure}
\centerline{%
\vbox to 2.4in{\vss
   \hbox to 3.3in{\includegraphics{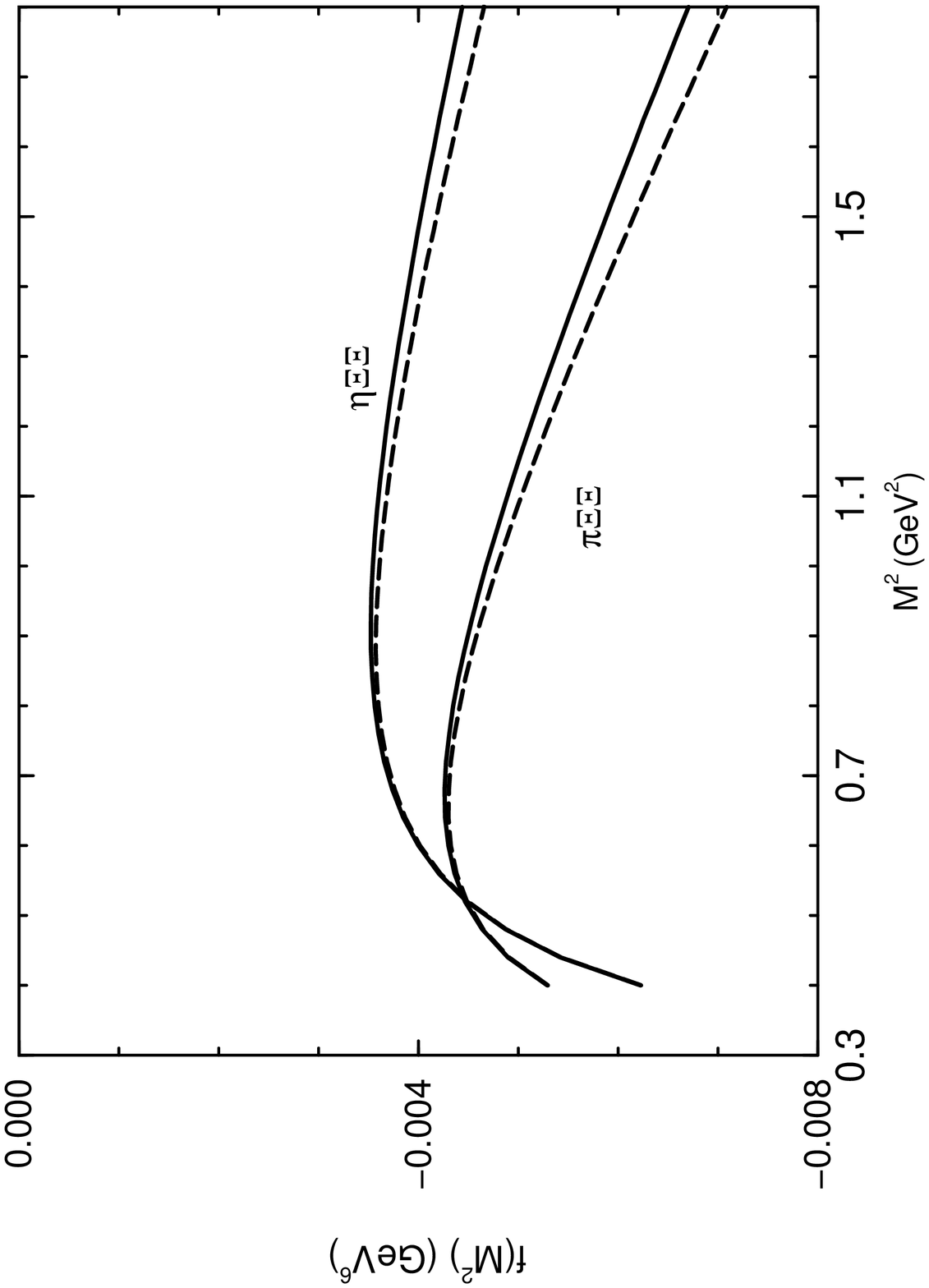}\hss}}
}
\eject
\end{figure}

\begin{figure}
\caption{ The Borel curves for $\pi \Sigma \Sigma$ and
$\eta \Sigma \Sigma$ in the SU(3) limit.  In the
curve for $\pi \Sigma \Sigma$, there is no
sensitivity to the continuum threshold as the
OPE has only the power corrections.}
\label{fig3}
\end{figure}

\setlength{\textwidth}{6.1in}   
\setlength{\textheight}{9.in}  
\begin{figure}
\centerline{%
\vbox to 2.4in{\vss
   \hbox to 3.3in{\includegraphics{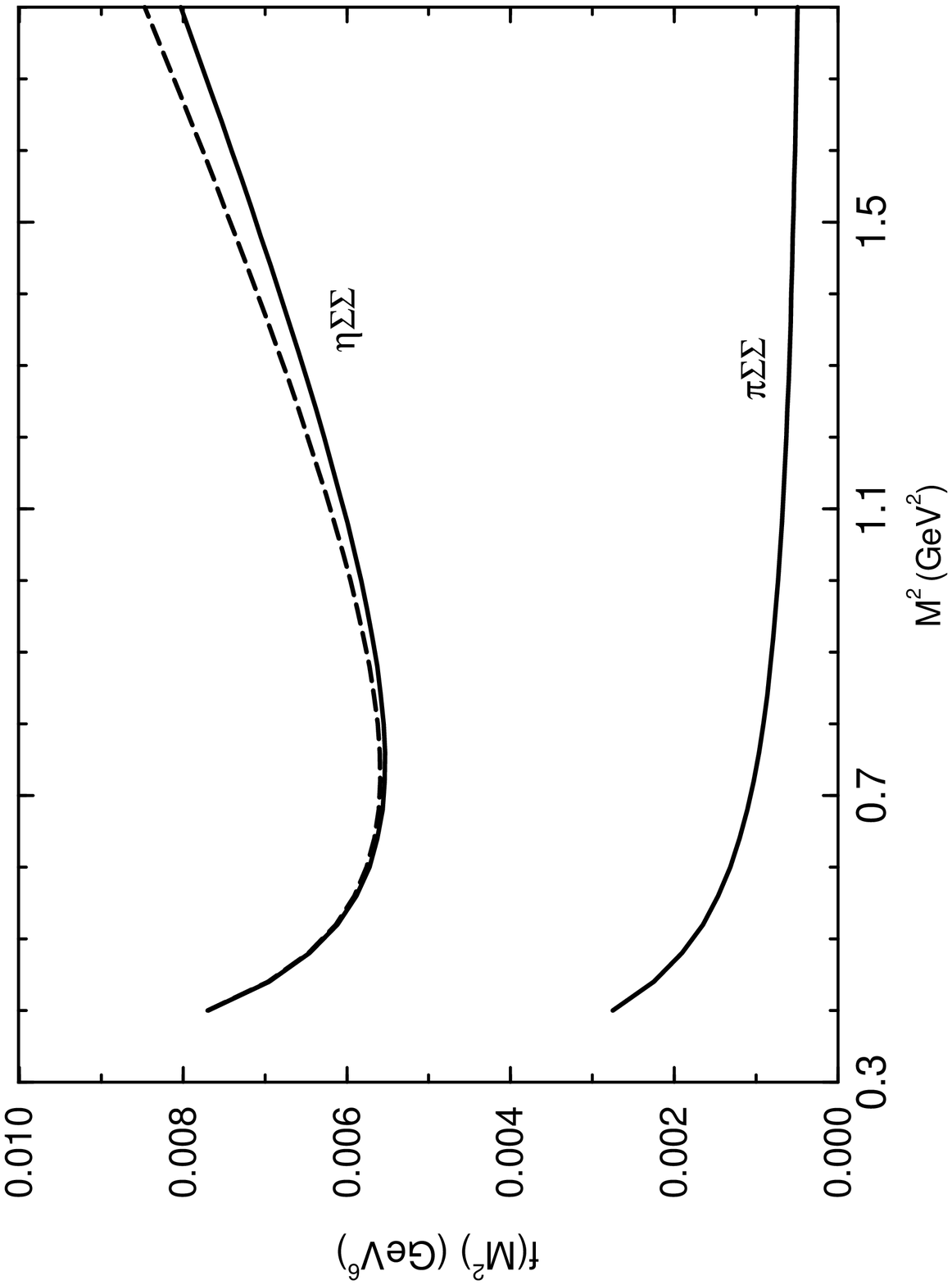}\hss}}
}
\vspace{100pt}
\end{figure}

\begin{figure}
\caption{ The Borel curve for 
$\eta NN$ beyond the SU(3) limit. The $\pi NN$
Borel curve is also shown for comparison. }
\label{fig4}
\end{figure}

\setlength{\textwidth}{6.1in}   
\setlength{\textheight}{9.in}  
\begin{figure}
\centerline{%
\vbox to 2.4in{\vss
   \hbox to 3.3in{\includegraphics{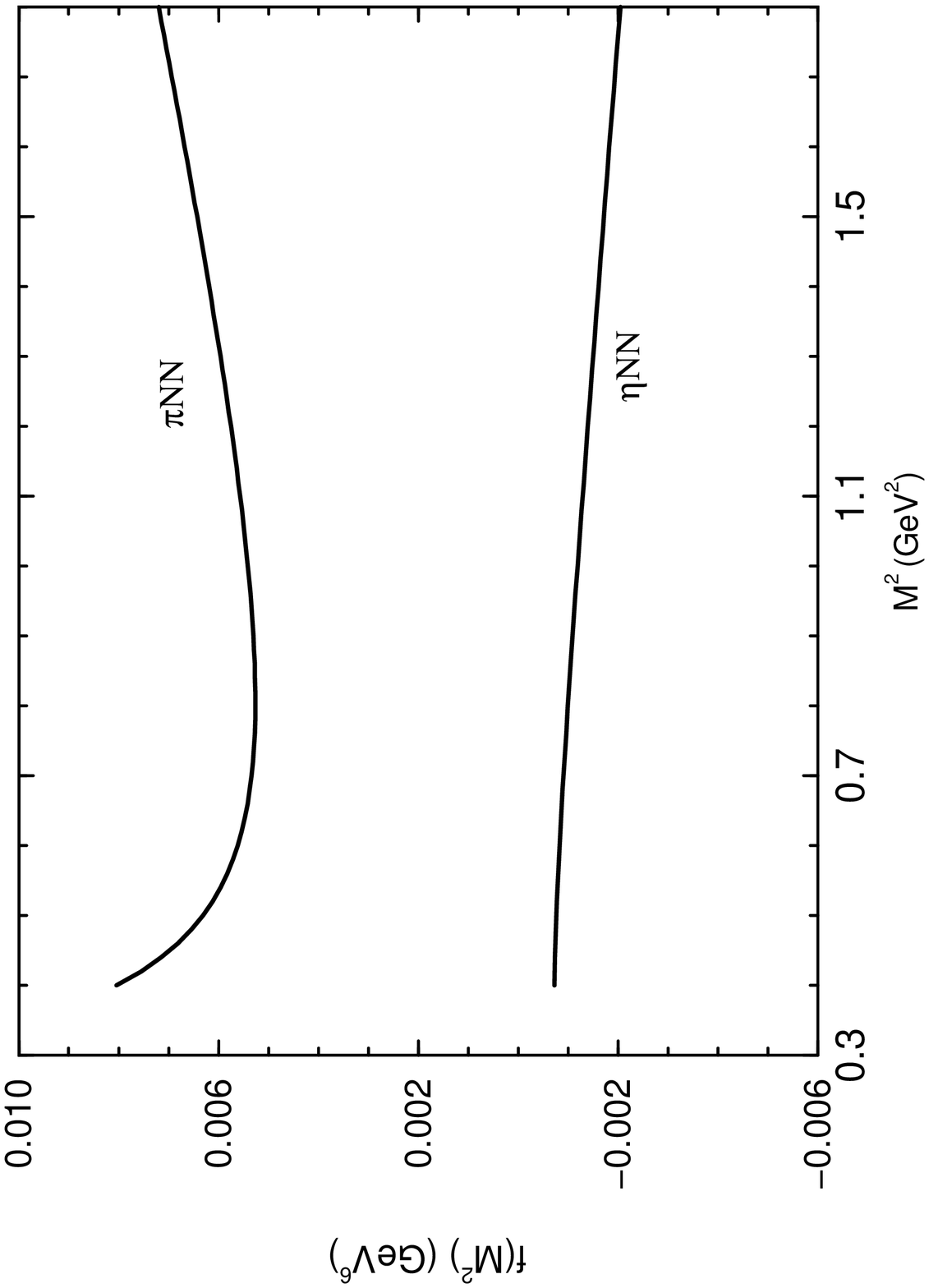}\hss}}
}
\eject
\end{figure}

\begin{figure}
\caption{ The Borel curves for $\pi \Xi \Xi$ and 
$\eta \Xi \Xi$ beyond the SU(3) limit. The continuum
threshold $S_0 = 3$ GeV$^2$ is used. Note, compared with
the SU(3) symmetric case in Fig.(3), the scale in
y-axis is much larger here.}
\label{fig5}
\end{figure}

\setlength{\textwidth}{6.1in}   
\setlength{\textheight}{9.in}  
\begin{figure}
\centerline{%
\vbox to 2.4in{\vss
   \hbox to 3.3in{\includegraphics{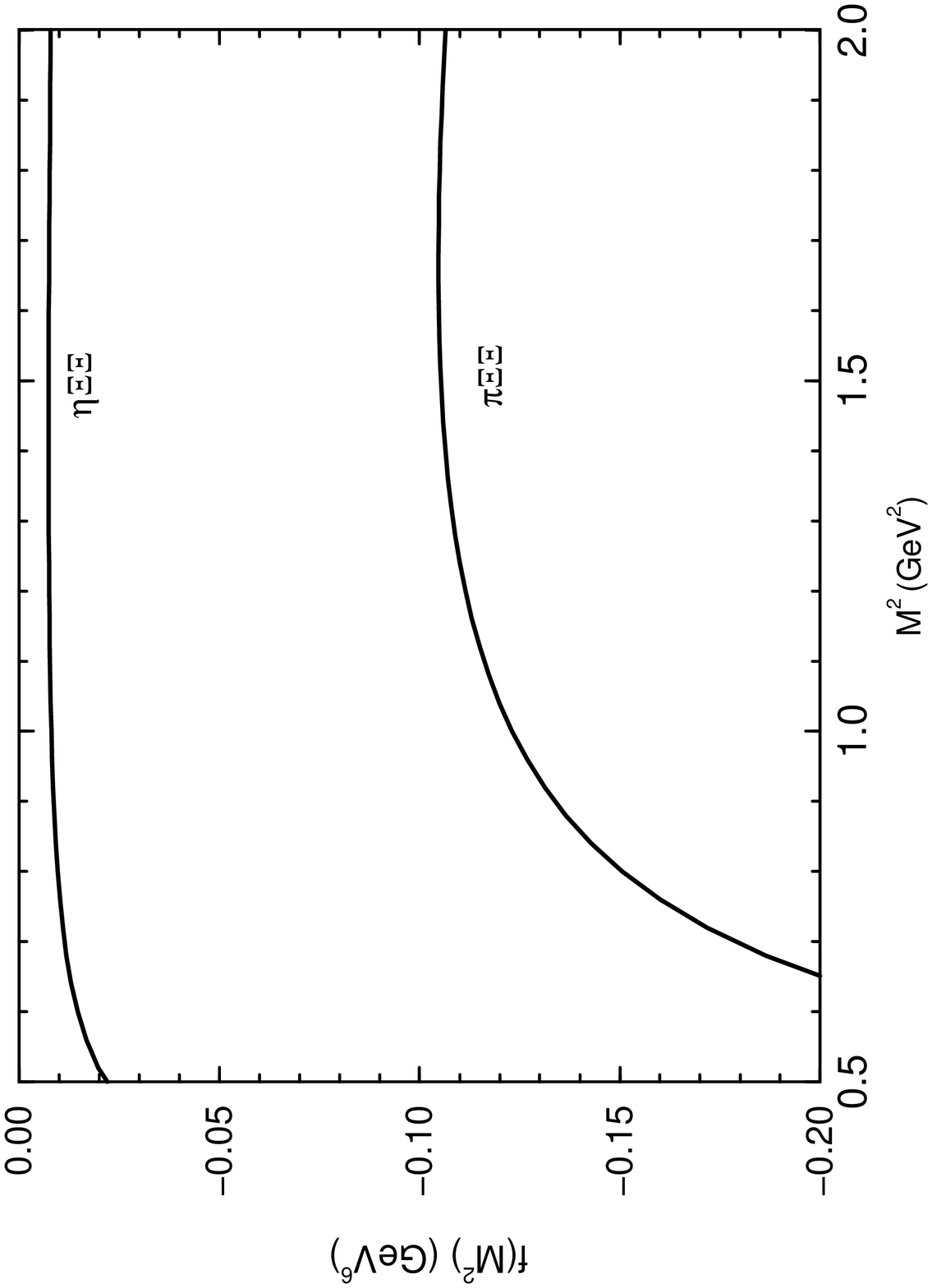}\hss}}
}
\vspace{100pt}
\end{figure}

\begin{figure}
\caption{ The Borel curves for $\pi \Sigma \Sigma$ and 
$\eta \Sigma \Sigma$ beyond the SU(3) limit. The continuum
threshold $S_0 = 3$ GeV$^2$ is used.}
\label{fig6}
\end{figure}

\setlength{\textwidth}{6.1in}   
\setlength{\textheight}{9.in}  
\begin{figure}
\centerline{%
\vbox to 2.4in{\vss
   \hbox to 3.3in{\includegraphics{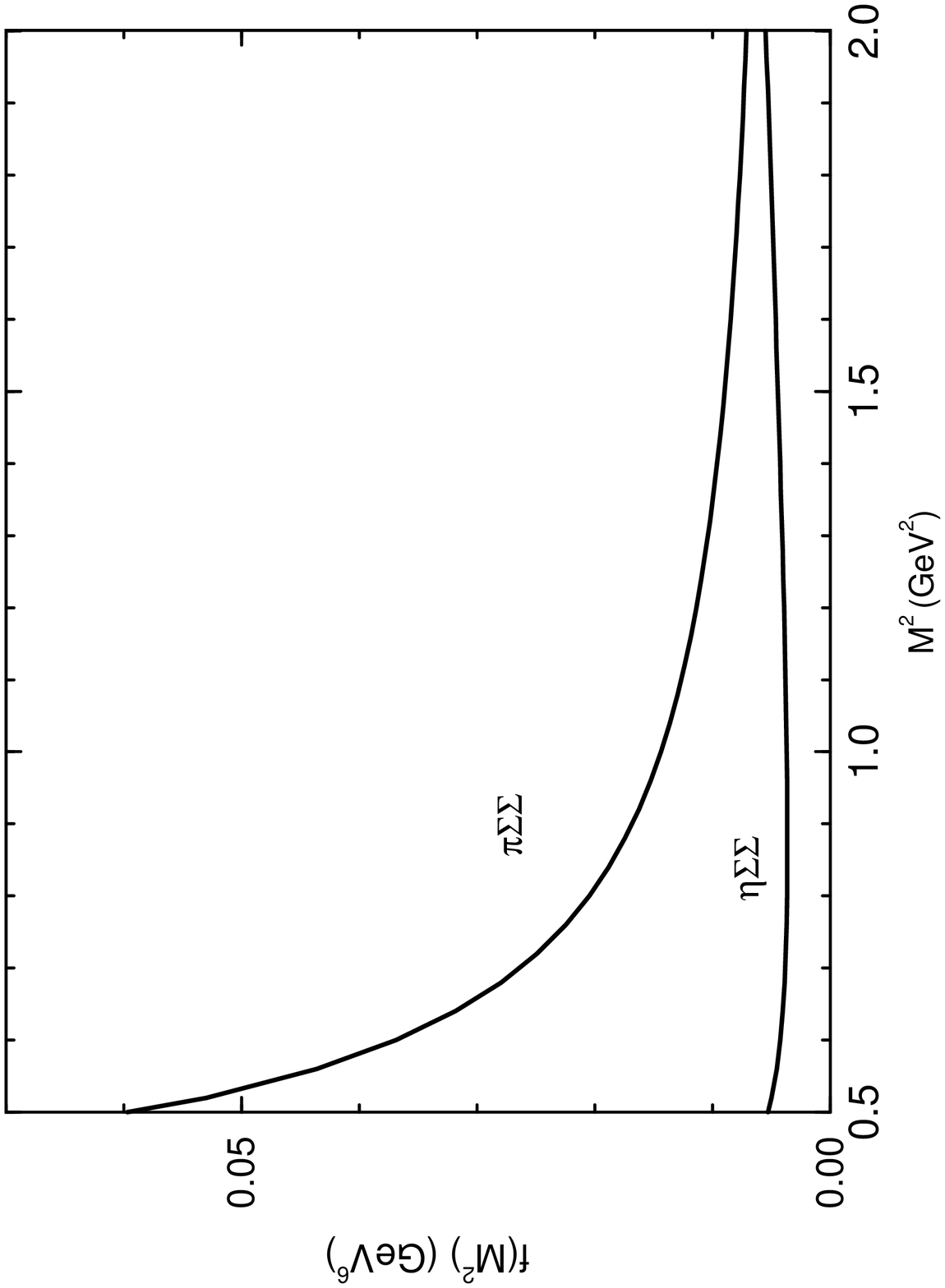}\hss}}
}
\end{figure}


\begin{references}
\bibitem{bonn}       {R. Machleidt in
                          {\it Advances in Nuclear Physics},
                          edited by J.W. Negele and E. Vogt
                          (Plenum, New York, 1989), Vol. 19.}
\bibitem{rijken}    {V. G. J. Stoks and Th. A. Rijken,
                            Phys. Rev. C {\bf 59}, 3009 (1999).;
                    Th. A. Rijken, V. G. J. Stoks and Y. Yamamoto,
                    Phys. Rev. C {\bf 59}, 21 (1999).}
\bibitem{swart}     {J. J. de Swart, 
                    Rev. Mod. Phys. {\bf 35}, 916 (1963).; {\bf 37}, 
                    326 (E) (1965).}
\bibitem{SVZ}     {M.A. Shifman, A.I. Vainshtein, and V.I. Zakharov,
                            Nucl. Phys. {\bf B 147}, 385, 448 (1979).}
\bibitem{qsr}     { L.J. Reinders, H. Rubinstein and S. Yazaki, 
                   Phys. Rep. {\bf 127}, 1 (1985).}
\bibitem{hung2}     {Hungchong Kim, Su Houng Lee and Makoto Oka,
                    Phys. Rev. D{\bf 60}, 034007 (1999).}
\bibitem{hung4}     {Hungchong Kim, nucl-th/9906081.}

\bibitem{hat}     {H. Shiomi and T. Hatsuda,
                            Nucl. Phys. {\bf A 594}, 294 (1995).}
\bibitem{hung1}     {Hungchong Kim, Su Houng Lee and Makoto Oka,
                    Phys. Lett. B{\bf 453} (1999) 199.}
\bibitem{hung3}     {Hungchong Kim, nucl-th/9904049.}
\bibitem{yazaki}     {L.J. Reinders, H. Rubinstein and S. Yazaki,
                            Nucl. Phys. {\bf B 213}, 109 (1983).;
                     T. Meissner and E. M. Henley,
                            Phys. Rev. C {\bf 55}, 3093 (1997).}
\bibitem{maltman}     {K. Maltman,
                            Phys. Rev. C {\bf 57}, 69 (1998).}
\bibitem{ioffe1}     {B. L. Ioffe,
                            Nucl. Phys.  {\bf B188}, 317 (1981).}
\bibitem{ioffe2}     {B. L. Ioffe and A. V. Smilga,
                            Nucl. Phys. {\bf B 232}, 109 (1984).}
 
\bibitem{bely}     {V. M. Belyaev, V. M. Braun, A. Khodjamirian and R. R\"uckl,
                            Phys. Rev. D {\bf 51}, 6177 (1995).}
\bibitem{wilson}     {J. Pasupathy, J. P. Singh, S. L. Wilson and C. B. Chiu,
                            Phys. Rev. D {\bf 36}, 1442 (1987).;
                      S. L. Wilson, Ph.D thesis, 
                      University of Texas at Austin, 1987.}
\bibitem{braun}     {V. M. Braun and I. B. Filyanov,
                            Z. Phys. C{\bf 48}, 239 (1990).}
\bibitem{zhit}       {A. R. Zhitnitsky, I. R. Zhitnitsky and V. L. Chernyak,
                            Yad. Fiz. {\bf 41}, 445 (1985).}
\bibitem{ioffe3}     {B. L. Ioffe,
                            Phys. At. Nucl. {\bf 58}, 1408 (1995).}
\bibitem{rat}     {P. G. Ratcliffe,
                    Phys. Lett. B{\bf 365}, 383 (1996).; 
                    hep-ph/9710458.}
\bibitem{malt2}     {K. Maltman,
                            hep-ph/9904370.}
\bibitem{antonio}     {A. Pich and J. Prades,
                            hep-ph/9909244.}
\bibitem{nemoto}    {M. Takizawa, Y. Nemoto and M. Oka,
                            Phys. Rev. D {\bf 55}, 4083 (1997).}
\bibitem{dey}     {J. Dey,  M. Dey, M. S. Roy,
                    Phys. Lett. B{\bf 443}, 293 (1998).; 
                   J. Dey,  M. Dey, T. Frederico, L. Tomio,
                   Mod. Phys. Lett. A {\bf 12}, 2193 (1997).}
\bibitem{chiu}     {C. B. Chiu,J. Pasupathy and S. L. Wilson,
                            Phys. Rev. D {\bf 32}, 1786 (1985).}
\bibitem{weber}     {M. Kirchbach and H. J. Weber,
                            Comments Nucl. Part. Phys.{\bf 22} 171 (1998).}
\end{references}
\end{document}